\documentclass[floats,aip,cha,reprint,superscriptaddress]{revtex4-1}

\usepackage[colorlinks,bookmarks=false,citecolor=blue,linkcolor=blue,urlcolor=blue]{hyperref}
\usepackage[all]{hypcap}   

\usepackage{amsmath,amssymb}
\usepackage{graphicx}
\graphicspath{{figures/}{/}}

\usepackage{verbatim}
\usepackage{color}

\usepackage{placeins}    
\usepackage{flafter}     

\usepackage{bm}

\usepackage{prettyref}
\newrefformat{sec}{Sec.~\ref{#1}}
\newrefformat{app}{Appendix~\ref{#1}}
\newrefformat{fig}{Fig.~\ref{#1}}
\newrefformat{tab}{Tab.~\ref{#1}}
\newrefformat{eq}{Eq.~(\ref{#1})}

\newcommand{\zeroD}{\textsc{0d}}
\newcommand{\oneD}{\textsc{1d}}
\newcommand{\twoD}{\textsc{2d}}
\newcommand{\threeD}{\textsc{3d}}
\newcommand{\fourD}{\textsc{4d}}

\newcommand{\ud}{\mathrm{d}}

\newcommand{\NHIM}{\textsc{NHIM}}
\newcommand{\NHIMs}{\textsc{NHIMs}}

\newcommand{\nhimmani}{\ensuremath{\bm X}}

\newcommand{\nhimpomn}[2]{\ensuremath{\nhimmani^{{#1}}_{{#2}} }}
\newcommand{\nhimpomnmin}[1]{\ensuremath{\nhimmani_{\text{min}}^{{#1}} }}
\newcommand{\nhimpomnmax}[1]{\ensuremath{\nhimmani_{\text{max}}^{{#1}} }}
\newcommand{\nhimpomnmini}[2]{\ensuremath{\nhimmani_{\text{min}, {#2}}^{{#1}} }}
\newcommand{\nhimpomnmaxi}[2]{\ensuremath{\nhimmani_{\text{max}, {#2}}^{{#1}} }}

\newcommand{\manistable}{\ensuremath{W^{\text{s}}}}
\newcommand{\maniunstable}{\ensuremath{W^{\text{u}}}}

\newcommand{\mappingtwod}{\ensuremath{f_{\twoD}}}

\newcommand{\mappingfourd}{\ensuremath{f_{\fourD}}}
\newcommand{\mapping}{\ensuremath{f}}

\newcommand{\minusoneform}{\ensuremath{\alpha}}
\newcommand{\gentwoform}{\ensuremath{\lambda}}
\newcommand{\gentwoformtwod}{\ensuremath{\lambda_\textsc{2d}}}
\newcommand{\gentwoformfourd}{\ensuremath{\lambda_\textsc{4d}}}

\newcommand{\invariantset}{\ensuremath{B}}

\newcommand{\regionone}{\ensuremath{\text{I}}}
\newcommand{\regiontwo}{\ensuremath{\text{II}}}

\newcommand{\flux}{\ensuremath{\Phi}}

\newcommand{\pomn}[1]{\ensuremath{\bm x^{{#1}} }}
\newcommand{\pomni}[2]{\ensuremath{\bm x^{{#1}}_{{#2}} }}
\newcommand{\pomnmin}[1]{\ensuremath{\pomn{#1}_{\text{min} }}}
\newcommand{\pomnmini}[2]{\ensuremath{\pomn{#1}_{\text{min},{#2}}}}
\newcommand{\pomnmax}[1]{\ensuremath{\pomn{#1}_{\text{max} }}}
\newcommand{\pomnmaxi}[2]{\ensuremath{\pomn{#1}_{\text{max}, {#2}}}}

\newcommand{\ktwod}{\ensuremath{K}}
\newcommand{\kcrit}{\ensuremath{K_c}}
\newcommand{\kone}{\ensuremath{K_1}}
\newcommand{\ktwo}{\ensuremath{K_2}}
\newcommand{\coup}{\ensuremath{\xi}}

\newcommand{\cospi}[1]{\ensuremath{\cos ( 2 \pi {#1} ) }}

\newcommand{\potargs}[3][]{\ensuremath{V_{#1} (#2, #3) }}

\newcommand{\HIDDEN}[1]{}

\newcommand{\uproman}[1]{\uppercase\expandafter{\romannumeral#1}}

\makeatletter
\let\Hy@backout\@gobble
\makeatother

\begin{document}

\title{Partial barriers to chaotic transport in
  4D symplectic maps}

\author{Markus Firmbach}
\affiliation{Technische Universit\"at Dresden,
 Institut f\"ur Theoretische Physik and Center for Dynamics,
 01062 Dresden, Germany}

\author{Arnd B\"acker}
\affiliation{Technische Universit\"at Dresden,
 Institut f\"ur Theoretische Physik and Center for Dynamics,
 01062 Dresden, Germany}

\author{Roland Ketzmerick}
\affiliation{Technische Universit\"at Dresden,
 Institut f\"ur Theoretische Physik and Center for Dynamics,
 01062 Dresden, Germany}

\date{\today}

\begin{abstract}
Chaotic transport in Hamiltonian systems is often restricted due to the presence of partial barriers,
leading to a limited flux between different regions in phase phase.
Typically, the most restrictive partial barrier in a \twoD{} symplectic map is based on a cantorus,
the Cantor set remnants of a broken \oneD{} torus.
For a \fourD{} symplectic map we establish
a partial barrier based on what we call a cantorus-\NHIM{},
a normally hyperbolic invariant manifold (\NHIM{}) with the structure of a cantorus.
Using a flux formula, we
determine the global \fourD{} flux across a partial barrier based on a cantorus-\NHIM{} by
approximating it with high-order periodic \NHIMs{}.
In addition, we introduce a local \threeD{} flux depending on the position along a resonance channel,
which is relevant in the presence of slow Arnold diffusion.
Moreover, for a partial barrier composed of stable and unstable manifolds of a \NHIM{}
we utilize periodic \NHIMs{} to quantify the corresponding flux.
\end{abstract}
\maketitle

\noindent
\begin{quotation}
The dynamics of Hamiltonian systems shows a large variety of behavior
ranging from integrable, over mixed to fully chaotic dynamics.
Transport of particles between different chaotic regions in phase space
can be very fast or completely inhibited.
In many situation slow transport occurs due to so-called partial barriers.
For lower-dimensional systems there exists a well-established analytical description
for the flux between regions separated by such partial-barriers.
However, for higher-dimensional systems it is not even clear which
partial barrier is most restrictive.
In this paper we provide for four-dimensional symplectic maps
such partial barriers, generalizing from the two-dimensional case.
The corresponding flux can be determined from a flux formula
using invariant sets of the map.
\end{quotation}

\section{Introduction}

Understanding chaotic transport
in the phase space of a Hamiltonian system is essential
in many areas of physics,
e.g., for the description of atoms and molecules
\cite{SchBuc2001, WigWieJafUze2001, TodKomKonBerRic2005, GekMaiBarUze2006, PasChaUze2008, WaaSchWig2008, ManKes2007, ManKes2014},
celestial motion and satellites
\cite{MurHol2001, Cin2002, DaqRosAleDelValRos2016, JafRosLoMarFarUze2002, DaqGkoRos2018, GkoDaqSkoTsiEft2019},
and particle accelerators  \cite{DumLas1993, VraIslBou1997, Pap2014}.
Transport in the chaotic component, however, is often impeded
due to the presence of partial barriers.
Such partial barriers in time-discrete maps
allow just for a small exchange of phase-space volume across them
\cite{KayMeiPer1984a, KayMeiPer1984b, Mei1992, Mei2015}.
In transition state theory \cite{TruGarKli1996, WaaWig2004, WaaSchWig2008, WaaWig2010, KayStr2014, EzrWig2018}
for time-continuous systems
they correspond to
dividing surfaces in phase space,
e.g.\ separating reactants and products of a
chemical reaction.

Here we focus on time-discrete symplectic maps.
Time-continuous Hamiltonian systems
can be reduced to time-discrete symplectic maps,
either for autonomous systems by considering energy conservation
and introducing a Poincar\'e section,
or for time-periodic systems by considering
stroboscopic sections \cite{LicLie1992}.
Partial barriers to chaotic transport
and the associated flux across them
are well understood
in two-dimensional (\twoD{}) symplectic maps \cite{KayMeiPer1984b, Mei1992, Mei2015}.
Many systems, however, have more degrees of freedom
and thus a higher-dimensional phase space
in which a similar understanding of partial barriers is still lacking.

In \twoD{} symplectic maps~\cite{Mei1992} one has \oneD{} invariant tori.
They are of codimension one and due to their invariance form
absolute barriers to chaotic transport,
separating different regions in the \twoD{} phase space.
Partial barriers are also \oneD{} closed curves,
but in contrast to tori not invariant under the map and therefore
allow for a small flux across them.
Typically,
the most restrictive partial barrier is based on a cantorus
\cite{Per1979,KayMeiPer1984b},
which arises from a broken regular torus (of irrational frequency)
whose remnants form a Cantor set.
Properties of such a cantorus
are conveniently determined from pairs of \zeroD{} hyperbolic
periodic orbits in the limit of large periods.
Remarkably, the flux across the cantorus
can be determined by the pairs of periodic orbits entering an action formula
which converges for increasing period \cite{KayMeiPer1984b, Mei1992}.
There exists another class of partial barriers with a typically
less restrictive flux, which originate from a hyperbolic fixed point
and its \oneD{} stable and unstable manifolds.
Note that for \threeD{} volume-preserving maps partial barriers
have for example been studied in Refs.~\cite{LomMei2009, FoxLla2015,FoxMei2013,FoxMei2016, DasBae2020}.

In a \fourD{} symplectic map, which is the minimal
higher-dimensional extension of a \twoD{} symplectic map,
new interesting properties arise:
(i)
A regular torus is two-dimensional and thus of codimension two
in the \fourD{} phase space.
Therefore it is no absolute barrier to chaotic transport.
Consequently, there is just one globally connected chaotic component.
(ii)
Chaotic transport is organized along resonance channels
\cite{Chi1979,Ten1982,WooLicLie1990, LicLie1992, Hal1999,Cin2002,CinGioSim2003,HonKan2005,CinEftGioMes2014,OnkLanKetBae2016,LanBaeKet2016}
and is slow due to the phenomenon of
Arnold diffusion~\cite{Arn1964, Chi1979, Loc1999, Dum2014}.
(iii)
In a higher-dimensional phase space
there exists a new type of invariant set, called
normally hyperbolic invariant manifold (\NHIM{})
\cite{Fen1972, HirPugShu1977, Man1978, Wig1994},
see also the introduction in Ref.~\cite{Eld2013}.
In the case of a \fourD{} map
a \NHIM{} is a \twoD{} manifold
and generalizes a \zeroD{} hyperbolic fixed point
of \twoD{} symplectic maps.
An important property of a \NHIM{} is its persistence
under (small) changes of parameters.

The existence of partial barriers restricting chaotic transport
in \fourD{} symplectic maps
is suggested by numerical observations, see e.g.\
\cite{KanBag1985, MarDavEzr1987, Las1993, LanBaeKet2016, FirLanKetBae2018}.
Orbits diffuse for a long time in a resonance channel
and once in a while
make a transition to another neighboring resonance channel,
see e.g.\ the supplementary video\cite{FirLanKetBae2018-vid-Fig7} corresponding to Fig.~7
in Ref.~\cite{FirLanKetBae2018}.
But in contrast to \twoD{} symplectic maps,
it is not clear
which partial barriers are of relevance to chaotic transport,
how to construct them, and how to compute the flux across them.

Partial barriers need to have codimension one to separate
regions in phase space. For a \fourD{} map they are \threeD{} manifolds.
Therefore the
\threeD{} stable and unstable manifolds of
a \twoD{} \NHIM{} naturally combine to a partial barrier
\cite{Wig1990, GilEzr1991}.
In \twoD{} maps typically
the most restrictive partial barrier
is based on a cantorus and
one may wonder whether this carries over to \fourD{} maps.
While cantori in higher-dimensional maps have been studied, see e.g.\
\cite{CheMacMei1990, BolMei1993, FoxMei2013},
they are not of sufficient codimension to provide a partial barrier
in \fourD{} maps.
Is there any other higher-dimensional generalization of a cantorus
leading to a restrictive partial barrier?

In this paper we introduce partial barriers based on
what we call a cantorus-\NHIM{}.
This invariant object is in the uncoupled limit
the Cartesian product of a cantorus of one \twoD{} map
and the entire phase space of a second \twoD{} map.
It can be approximated by periodic \NHIMs{}
which consist of several \twoD{} manifolds.
We determine the \fourD{} flux across a partial barrier
based on pairs of periodic \NHIMs{}.
This flux is determined from
the \twoD{} periodic \NHIMs{} only.
It is thus independent of the specific partial barrier.
It converges for increasing periods $n$
when an irrational frequency is approached,
giving the \fourD{} flux across the cantorus-\NHIM{}.
To our knowledge, the partial barrier based on the cantorus-\NHIM{}
is the first explicit realization of a
partial barrier associated with an irrational frequency line,
as proposed in Ref.~\cite{MarDavEzr1987}.

Furthermore, we argue that in the presence of slow Arnold diffusion
along a resonance channel the above global \fourD{} flux is not relevant.
In contrast, transport to a neighboring resonance channel
is described by a local \threeD{} flux,
which depends on the position along the resonance channel.
We capture the essential properties of the local flux
by a contribution where only periodic \NHIMs{} enter.

Moreover, we study a partial barrier composed of the \threeD{}
stable and unstable manifolds of a \NHIM{}.
We determine its global and local
flux using periodic \NHIMs{}.
We demonstrate these results for the case of the \fourD{} standard map.
An extension to time-continuous systems and transition state theory seems possible.

The paper is organized as follows.
In \prettyref{sec:pbfluxgen} we briefly review the
concept of a partial barrier in an $n$-dimensional phase space
and recall how the flux across a partial barrier
can be determined from invariant sets of the map.
In  \prettyref{sec:pbflux2d} we
apply the general concepts to the \twoD{} standard map
and review how the flux across a cantorus is approximated using periodic orbits.
In \prettyref{sec:flux4d} the application to
the \fourD{} standard map for a cantorus-\NHIM{}
is presented.
In particular, we introduce periodic \NHIMs{} and determine the
flux across partial barriers based on them.
By considering a sequence of periodic \NHIMs{} we approximate
a cantorus-\NHIM{} and the flux across it.
In \prettyref{sec:localflux} we
introduce the local flux and capture its essential properties
by a contribution where only periodic \NHIMs{} enter.
Furthermore in \prettyref{sec:pbmanifold},
we study the global and local
flux across a partial barrier which is composed of the
stable and unstable manifold of a \NHIM{}.
Section \ref{sec:outlook} gives both a summary of the results
and an outlook to future research.
In \prettyref{app:dynonnhim} we depict the
phase-space portraits of the non-integrable dynamics on the periodic \NHIMs{}
for increasing periods.
In \prettyref{app:lobearea} and \ref{app:nhimcont}
we give the derivation for flux formulas
used in the main text.


\section{Partial barriers and flux}\label{sec:pbfluxgen}

\begin{figure}
	\includegraphics{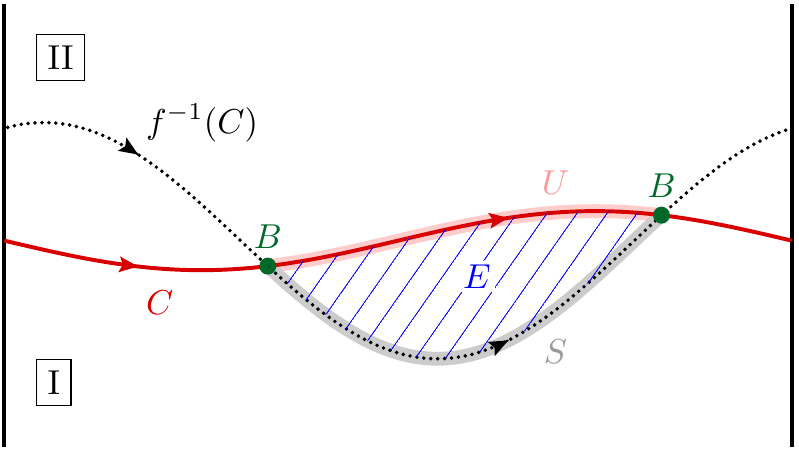}
	\caption{
		Partial barrier $C$ (red solid line) in a \twoD{} cylindrical phase space
		and its preimage $\mapping^{-1} (C)$ (black dotted line)
		confining the exit lobe $E$ (blue hatched)
		that is transported from region~\regionone{} to region~\regiontwo{}
		under the map \mapping.
		The oriented boundary of $E$ is given by $\partial E = U - S$.
		The invariant points $B$ (dark green points) are the boundary of $U$
		and also of $S$.
	}\label{fig:sketch_lobe}
\end{figure}

In the following we shortly review the notion of
partial barriers and flux for a
volume-preserving map \mapping{} acting on
an $n$-dimensional phase space,
see Ref.~\cite{Mei2015} for an extended exposition.
This will lead to the flux formula, \prettyref{eq:flux_general},
which will be applied to \twoD{} and \fourD{} maps in this paper.

A closed manifold $C$ of codimension one, i.e.\ with $(n-1)$ dimensions,
divides phase space into two distinct regions,
in the following referred to as region~\regionone{} and region~\regiontwo{}.
This is illustrated in
\prettyref{fig:sketch_lobe}
for $n=2$ and a \oneD{} manifold~$C$.
If the manifold $C$ is invariant under the map \mapping{},
i.e.\ it is identical to its preimage $\mapping{}^{-1}(C)$,
it is a barrier to transport.
If $C$ is not invariant, it
acts as a \emph{partial barrier},
since it allows for an exchange of phase-space
volume from region~\regionone{} into~\regiontwo{},
and vice versa.

The \emph{flux} \flux{} is the
volume of phase space which is transported
across the partial barrier,
e.g.\ from region~\regionone{} to~\regiontwo{},
 upon each iteration of the map~\mapping{}.
The flux~\flux{} can be determined from the partial barrier $C$
and its preimage $\mapping^{-1} (C)$,
see \prettyref{fig:sketch_lobe}:
Together both manifolds $C$ and \mapping{}${}^{-1}(C)$
confine (several) $n$-dimensional volumes, denoted as \emph{lobes},
located on opposite sides of the partial barrier $C$.
The lobes located in region~\regionone{}
are the exit set $E$ of region~\regionone{}, as points in $E$ are
transported across $C$ into region~\regiontwo{} under one iteration of the map \mapping{}.
Thus the flux \flux{} across the partial barrier
is given by the volume of $E$.
Conversely, the lobes located in region~\regiontwo{}
are the exit set of region~\regiontwo{}, as points in these lobes
are transported across $C$ into region~\regionone{} under one iteration
of the map \mapping{}.
We will restrict to the case of zero net flux across the partial barrier $C$,
which holds for exact volume-preserving maps~\cite{Mei2015},
i.e.\ the flux~\flux{}
also describes the identical phase-space volume
transported from region
\regiontwo{} to~\regionone{}.
The notion of a turnstile describes the combined action of the
exit lobe of region~\regionone{} and the
exit lobe of region~\regiontwo{}
under iteration of the map \mapping,
which interchange phase-space volume
between these regions like a rotating door
\cite{KayMeiPer1984a, Mei2015}.

The flux \flux{} across a partial barrier will be determined using differential forms,
which allow for a convenient description independent of the number $n$ of
phase-space dimensions~\cite{Eas1991, LomMei2009, Mei2015},
\begin{equation}
	\flux
	= \int \limits_{E} \Omega
	= \int \limits_{E} \ud\minusoneform
	= \int \limits_{\partial E} \minusoneform
	= \int \limits_{U} \minusoneform
	- \int \limits_{S} \minusoneform
	,
	\label{eq:flux_1}
\end{equation}
where $\Omega$ is a volume form
that we assume to be exact,
i.e.\ $\Omega = \ud\minusoneform$,
with $\minusoneform$ being some $(n-1)$-form.
Stokes theorem is used in \prettyref{eq:flux_1},
such that the dimension of the integral
is reduced by one.
In the last step of
\prettyref{eq:flux_1}
the boundary $\partial E$ of $E$ is decomposed,
$\partial E = U - S$,
into an oriented
subset $U$ of the partial barrier,
$U \subset C$,
and an oriented
subset $S$ of the preimage,
$S \subset \mapping^{-1} (C)$,
see \prettyref{fig:sketch_lobe}.
The sets $U$ and $S$ are $(n-1)$-dimensional manifolds.

The restriction to an exact volume-preserving
map \mapping{} with
an $(n-1)$-form \minusoneform{},
allows to write~\cite{LomMei2009, Mei2015}
\begin{equation}
	\mapping^{*} \minusoneform - \minusoneform = \ud \gentwoform
	,
	\label{eq:volprev}
\end{equation}
where $\mapping^{*} \minusoneform$ is the pullback and
\gentwoform{} is an $(n-2)$ form
which is called the generating Lagrangian form.
Using this for $\minusoneform$ in the last integral of
\prettyref{eq:flux_1} gives
\begin{equation}
	\flux
	= \int \limits_{U} \minusoneform - \int \limits_{S} (\mapping^{*} \minusoneform - \ud \gentwoform)
	.
	\label{eq:flux_2}
\end{equation}
By applying Stokes theorem to the last term,
using $\partial S = \partial U$,
and by using
$\int \limits_{S} \mapping^{*} \minusoneform = \int \limits_{\mapping(S)} \minusoneform$
one gets
\begin{equation}
	\flux
	= \int \limits_{\partial U} \gentwoform
	+ \int \limits_{U} \minusoneform - \int \limits_{\mapping(S)} \minusoneform
	.
	\label{eq:flux_3}
\end{equation}
Note that the iterated set $\mapping(S)$
as well as $U$ are subsets of the partial barrier $C$.
The $(n-2)$-dimensional boundary $\partial U$ of $U$ in
\prettyref{eq:flux_3}
is also the boundary $\partial S$ of $S$.
It is given by the intersection of the partial barrier and its preimage,
$\partial U = \partial S = C \cap \mapping^{-1} (C)$,
see \prettyref{fig:sketch_lobe}.
In general one has several exit lobes of region~\regionone{}
and then the sets $U$ and $S$
are composed of several pieces, which confine all exit lobes.

In this paper we consider $(n-1)$-dimensional
partial barriers, which connect the
$(n-2)$-dimensional elements of an invariant set \invariantset{}.
Therefore the preimage of the partial barrier
contains the same invariant set  \invariantset{},
see \prettyref{fig:sketch_lobe}.
Thus the invariant set \invariantset{} is a subset of
the intersection of
the partial barrier and its preimage,
and thus $\invariantset \subset \partial U$.
Additionally, we require that there are no further intersections,
leading to
$\partial U = \invariantset$.
Note that this corresponds to \emph{hypothesis (H2)} in Ref.~\cite{Mei2015}.

Furthermore,
for this class of partial barriers,
the iterated set $\mapping(S) \subset U$ appearing in
\prettyref{eq:flux_3}
is, just like $S$, bounded by the invariant set
\invariantset{}.
Therefore $\mapping(S)$ is in fact identical to the subset $U$.
With $\mapping(S) = U$ the last two integrals in
\prettyref{eq:flux_3}
cancel, leading to~\cite{LomMei2009, Mei2015}
\begin{equation}
	\flux
	= \int \limits_{\invariantset} \gentwoform
	.
	\label{eq:flux_general}
\end{equation}
In the examples that will be studied in this paper,
the invariant set \invariantset{}
consists of two parts which are each invariant.
They occur at different ends of $U$ (for each lobe) and
thus contribute with different signs to
$\invariantset{} = \partial U$.

Remarkably, according to \prettyref{eq:flux_general},
the flux \flux{}
across this class of partial barriers
can be determined from the
invariant set \invariantset{} only.
As a consequence,
the flux \flux{} is independent of the specific choice of the partial barrier
as long as it contains the invariant set
\invariantset{}
and has no further intersections with its preimage.
It allows for determining the
$n$-dimensional flux
by an $(n-2)$-dimensional integral over
the invariant set \invariantset{}.

\section{Partial barriers and flux in \twoD{} maps}\label{sec:pbflux2d}
In this section we apply the
general formalism of partial barriers and flux,
in particular \prettyref{eq:flux_general},
to symplectic \twoD{} maps.
By reviewing some aspects of the well-established
theory of transport in \twoD{} maps \cite{KayMeiPer1984b, Mei1992, Mei2015}
we introduce some notation
needed for the generalization to symplectic \fourD{} maps in \prettyref{sec:flux4d}.
In particular, we consider partial barriers which are
based on periodic orbits,
as they allow to approximate the properties of a cantorus.
Cantori give rise to the most restrictive partial
barriers in \twoD{} maps.

\subsection{System}
\begin{figure}
    \includegraphics{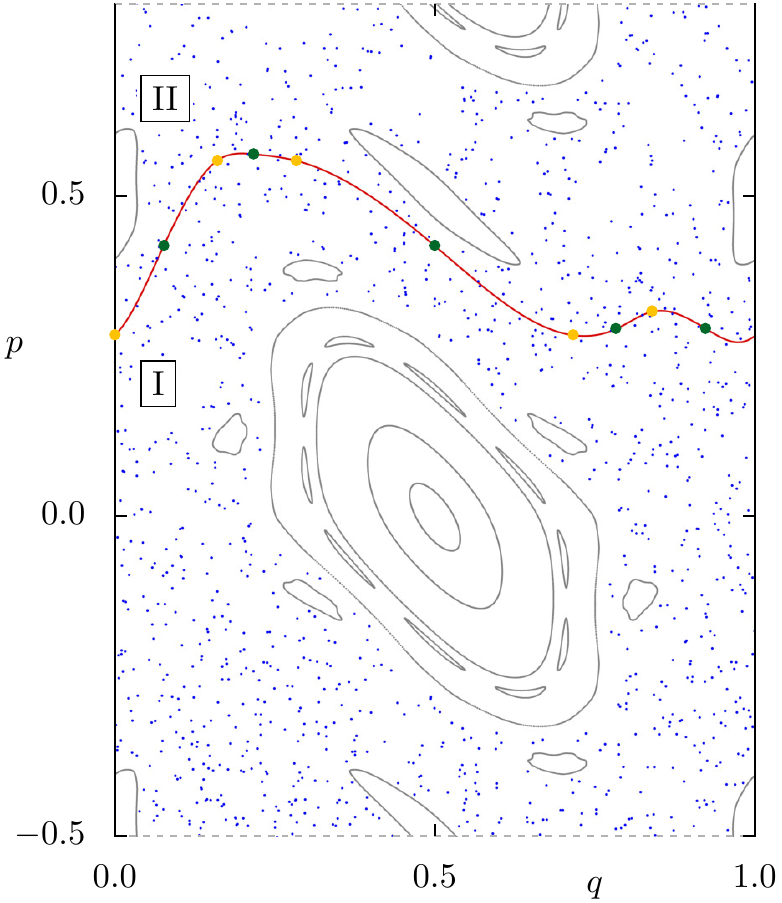}
    \caption{Phase space of the \twoD{} standard map \mappingtwod{} for $\ktwod=1.75$
        with regular tori (gray lines) and one chaotic orbit (blue points).
        A partial barrier (red solid line) separates region~$\regionone$ and~\regiontwo{}.
        It is constructed by alternatingly joining the points of
        the periodic orbits \pomnmax{2:5} (dark green points)
        and \pomnmin{2:5}
        (light orange points).
    }\label{fig:ps2d}
\end{figure}

As a paradigmatic example of a \twoD{} map,  we consider the standard map\cite{Chi1979} \mappingtwod{},
\begin{equation}
\begin{aligned}
q' &= q + p \\
p' &= p - \frac{\ud V}{\ud q}(q')
\end{aligned}
\label{eq:map_std2d}
\end{equation}
on the cylinder $(q, p) \in [0,1) \times \mathbb{R}$
with periodic boundary conditions imposed in $q$
and kicking potential
$V(q) = \frac{\ktwod}{4\pi^2} \cos(2\pi q)$.

At kicking strength $\ktwod=0$ the mapping is integrable
and the dynamics is foliated into rotational invariant \oneD{} tori
with fixed $p$.
For $\ktwod>0$ tori with rational frequency break-up according
to the Poincar\'e-Birkhoff theorem, while tori with sufficiently
irrational frequency persist according to the KAM theorem~\cite{LicLie1992}.
These curves of codimension one form absolute barriers
for chaotic transport in $p$-direction.
The last rotational invariant torus breaks up\cite{Gre1979}
at $\kcrit \approx 0.97$.
A mixed phase space for $\ktwod=1.75$ is shown in \prettyref{fig:ps2d}
where regions of regular and chaotic dynamics coexist.
The chaotic region extends in the $p$-direction
and transport is not
blocked by any rotational invariant \oneD{} torus.
However, chaotic transport along the $p$-direction is impeded by
the presence of \oneD{} partial barriers.
They are based on either periodic orbits or on cantori
as invariant sets \invariantset{},
as discussed below.

\subsection{Periodic orbits}
We consider periodic orbits
with frequency
$\nu = \frac{m}{n}$,
consisting of $n$
periodic points
$\pomni{m:n}{i} = (q^{m:n}_i, p^{m:n}_i)$,
$i = 0, \dots, n-1$.
Each point
$\pomni{m:n}{i}$
is mapped to the next point,
$\mappingtwod(\pomni{m:n}{i}) = \pomni{m:n}{(i+1)\mod n}$,
and thus a fixed point of the $n$-fold map,
$\mappingtwod^{n} (\pomni{m:n}{i}) =  \pomni{m:n}{i}$.
The index $m$ describes the number of cycles in $q$
after $n$ iterations.
We denote the set of periodic points by
\begin{equation}
	\pomn{m:n} = \{ \pomni{m:n}{i} \; | \; i = 0, \dots, n-1\}
	,
\end{equation}
which is invariant
under one iteration of the map,
\begin{equation}
	\mappingtwod (\pomn{m:n}) =  \pomn{m:n}
	.
\end{equation}
These periodic orbits come in pairs,
namely one so-called minimizing periodic orbit \pomnmin{m:n}
and one minimax periodic orbit \pomnmax{m:n}.
Their existence is guaranteed by Aubry-Mather theory \cite{Mei1992, Mei2015}.
For the considered kicking strength $\ktwod=1.75$ the periodic orbit
\pomnmin{2:5} is hyperbolic and \pomnmax{2:5} is inverse hyperbolic, see
\prettyref{fig:ps2d}.

\subsection{Partial barriers based on periodic orbits}
\begin{figure*}
    \includegraphics{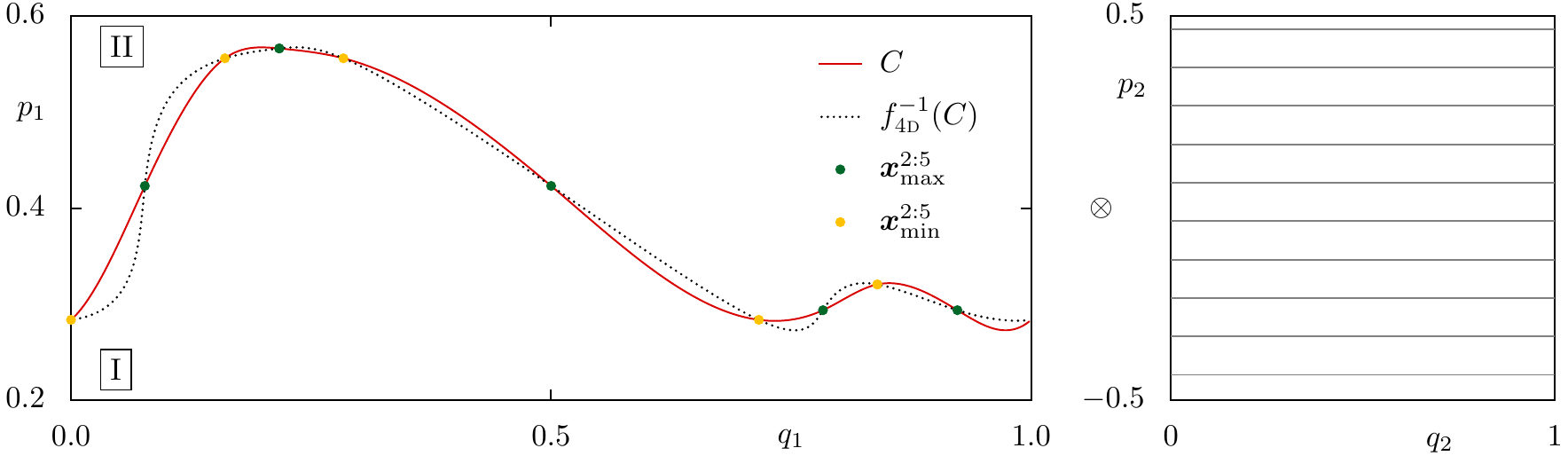}
    \caption{Product structure of the
        \fourD{} standard map \mappingfourd{}
        without coupling $(\coup = 0)$
        for $\kone=1.75$, $\ktwo=0$.
        Periodic orbits \pomnmax{2:5} (dark green points)
        and \pomnmin{2:5} (light orange points)
        are shown in $(q_1, p_1)$ space.
        A partial barrier $C$ (red solid line) connects these periodic points and
        separates region~\regionone{} from region~\regiontwo{}.
        Its preimage $\mappingfourd^{-1}(C)$ (black dotted line)
        intersects at the periodic points.
        Phase space in $(q_2, p_2)$ is integrable.
    }
    \label{fig:ps_product}
\end{figure*}

We consider partial barriers $C$ connecting points of the periodic orbits \pomnmin{m:n} and \pomnmax{m:n}
in an alternating manner.
One example of such a partial barrier $C$ is shown in \prettyref{fig:ps2d},
separating
region~\regionone{} from~\regiontwo{}. It connects points of the periodic orbits
\pomnmin{2:5} and \pomnmax{2:5}.
The preimage  \mapping${}^{-1} (C)$ intersects the partial barrier in the periodic points only,
as shown in \prettyref{fig:ps_product} (left).
The partial barrier and its preimage enclose $2n$ lobes,
which are alternatingly located above and below the partial barrier $C$.
All lobes in region~\regionone{} contribute to the flux~\flux{}
to region~\regiontwo{}.
Note that these criteria are fulfilled by infinitely many partial barriers,
which turn out to all have the same flux.

In order to determine the flux across partial barriers of this kind
we apply the flux formula, \prettyref{eq:flux_general}.
Since the \twoD{} standard map~\mappingtwod{} is exact area-preserving with
the one-form $\minusoneform = p \, \ud q$,
a generating zero-form (function) $\gentwoformtwod{}$ exists,
given by
\begin{equation}
\gentwoformtwod{} = \frac{1}{2} p^2 - V(q + p) ,
\label{eq:genform2d}
\end{equation}
and fulfilling \prettyref{eq:volprev}.
The invariant set $\invariantset{}$ in
the flux formula,
\prettyref{eq:flux_general},
consists of $\pomnmax{m:n}$ and $\pomnmin{m:n}$
at the ends of $U$,
contributing with different signs,
$\invariantset{} = \pomnmax{m:n} - \pomnmin{m:n}$.
This leads for the \twoD{} case to the
explicit expression
\cite{BenKad1984, KayMeiPer1984b, Mei1992, Mei2015}
\begin{equation}
\flux^{\twoD}
= \int \limits_{\invariantset} \gentwoformtwod
=  \sum_{i = 0}^{n-1} \: \gentwoformtwod (\pomnmaxi{m:n}{i})
\; - \;
\sum_{i = 0}^{n-1} \: \gentwoformtwod (\pomnmini{m:n}{i}) ,
\label{eq:flux_twod}
\end{equation}
where integration reduces to a summation of
the generating function evaluated on the
pair of periodic orbits.

Note that the same flux is found for a partial barrier with a different
construction giving just two lobes~\cite{Mei1992, Mei2015}:
One takes a piece of the above partial barrier,
e.g., from one point of \pomnmin{m:n} to the neighboring point of \pomnmin{m:n}
which goes through a point of \pomnmax{m:n}. This curve is then iterated backwards
$n-1$ times. Combining all these curves defines a partial barrier.
Its preimage, by construction, is identical to the partial barrier,
except for two lobes. The volume of the exit lobe,
see Eq.~(16) in Ref.~\cite{Mei2015},
leads to the same flux \prettyref{eq:flux_twod}.

\subsection{Cantorus}\label{subsec:cantorus2d}

A cantorus~\cite{Per1979,KayMeiPer1984b} emerges when a \oneD{} torus
with irrational frequency $\nu$ breaks up under parameter variation.
The remnants form a Cantor set.
Such a cantorus can be successively
approximated by the pair of periodic orbits,
\pomnmin{m:n} and \pomnmax{m:n},
with frequencies $\frac{m}{n}$ obtained by successive
truncations of the continued fraction expansion of its irrational
frequency $\nu$.

Partial barriers based on a
cantorus are most restrictive for chaotic transport in \twoD{} maps
with locally the smallest flux.
For the \twoD{} standard map \mappingtwod{}
the rotational invariant \oneD{} torus with
irrational frequency $\nu = 1- \sigma_{\text{G}} \approx 0.382$,
where $\sigma_{\text{G}}$ is the golden mean, persists longest
with increasing $\ktwod$.
It turns into a cantorus at the critical parameter
$\kcrit \approx 0.97$ and has globally the smallest flux for
$\ktwod > \kcrit$.

The flux~$\flux^{\twoD}$ of a partial barrier based on a cantorus
can be successively approximated by the flux
$\flux^{\twoD}$, \prettyref{eq:flux_twod},
of the pair of periodic orbits
\pomnmin{m:n} and \pomnmax{m:n} with increasing periods $n$.
Convergence of the flux
is shown for $\ktwod=1.75$ in \prettyref{fig:flux_4d_periods} and
\prettyref{tab:flux_4d} where the flux $\flux^\twoD$
is denoted as $\flux^{\fourD}(\coup = 0)$.

\section{Partial barriers and flux in \fourD{} maps}\label{sec:flux4d}
\subsection{Motivation}
In a higher-dimensional phase space much less is known about partial barriers.
A direct generalization from \twoD{} maps is not possible.
The reason is that invariant objects in phase space lack one or more dimensions in
order to constitute a (partial) barrier.
For example, in a \fourD{} map a regular torus is a \twoD{} manifold
and thus cannot separate the \fourD{} phase space into different regions.
Consequently, also a broken \twoD{} torus
cannot form a restrictive partial barrier.

The existence of restrictive partial barriers
in \fourD{} maps, however, is indicated by
numerical observations, see e.g.\ \cite{KanBag1985, MarDavEzr1987, Las1993, LanBaeKet2016, FirLanKetBae2018}.
Their origin and their construction is still an open question.

In the following we introduce a \threeD{} partial barrier
in a \fourD{} map
based on higher-dimensional generalizations of the periodic orbits
and cantori introduced in \prettyref{sec:pbflux2d}.
We will use the concept of
normally hyperbolic invariant manifolds
(\NHIMs{})~\cite{Fen1972, HirPugShu1977, Man1978, Wig1994, Eld2013}
to introduce a periodic \NHIM{} and a cantorus-\NHIM{}.
The flux across partial barriers based on these invariant sets
will be determined using
\prettyref{eq:flux_general}.

\subsection{System}\label{subsec:system}
We consider the
\fourD{} standard map\cite{Fro1972} \mappingfourd{},
\begin{align}
\begin{split}
q_1' &= q_1 + p_1 \\
q_2' &= q_2 + p_2 \\
p_1' &= p_1 - \potargs[1]{q_1'}{q_2'} \\
p_2' &= p_2 - \potargs[2]{q_1'}{q_2'},
\end{split}
\label{eq:map_stdfourd}
\end{align}
with the partial derivatives
$V_{i} = \frac{\partial V(q_1, q_2)}{\partial q_i}$
and potential
\begin{equation}
\begin{split}
V(q_1, q_2) = &\frac{\kone}{4\pi^2} \cospi{q_1} +
\frac{\ktwo}{4\pi^2} \cospi{q_2} \\
&+ \frac{\coup}{4\pi^2} \cospi{(q_1 + q_2)}
\end{split}
\label{eq:potential_stdfourd}
\end{equation}
composed of the potential of two \twoD{} standard maps
with kicking strengths \kone{} and \ktwo{}
and a coupling between both degrees of freedom of strength \coup{}.
We consider a cylinder in the first degree of freedom,
$(q_1, p_1) \in [0,1) \times \mathbb{R}$,
and a \twoD{} torus in the second,
$(q_2, p_2) \in [0,1) \times [-0.5, 0.5)$
with periodic boundary conditions imposed in $q_1, q_2, p_2$.

Since we aim at a higher-dimensional partial barrier
related to the cantorus of the \twoD{} map,
we fix $\kone = 1.75$.
Furthermore,
for the occurrence of \NHIMs{} (see discussion below)
we require the kicking strength $\ktwo$
of the second degree of freedom to be much smaller
than the kicking strength $\kone$
of the first degree of freedom, i.e.\ $\ktwo \ll \kone$.
For simplicity we set $\ktwo = 0$ throughout the paper.

If the coupling \coup{} is set to zero, i.e.\ $\coup = 0$,
the \fourD{} map \mappingfourd{}
is trivially related to the \twoD{} map \mappingtwod{}.
Phase-space structures are
given by the Cartesian product of the objects in
$(q_1, p_1)$ with the \twoD{} torus of the $(q_2, p_2)$ space.
This product structure is symbolically depicted in \prettyref{fig:ps_product}.
As function of $p_2$ one obtains a stack of identical \twoD{} maps in
the $(q_1, p_1)$ space with rotational dynamics in $q_2$ whose frequency
is given by $p_2$.
In particular, any \oneD{} partial barrier of
the \twoD{} map in the $(q_1, p_1)$ space
trivially extends to a \threeD{} partial barrier
for the \fourD{} map,
see Sec.~\ref{subsec:flux4d}.
Such a stack of systems has been used, e.g.,
in the famous Arnold model~\cite{Arn1964},
its \fourD{} map analogue~\cite{EasMeiRob2001},
and in higher-dimensional scattering\cite{JunMerSelZap2010,GonJun2012,GonDroJun2014}.

For non-zero but sufficiently small coupling strengths, ${\coup \ll 1}$,
one obtains a \fourD{} map
with topological features of the product structure for \mbox{$\coup=0$}.
This will be very helpful in visualizing
phase-space structures and quantifying the flux across partial barriers.

\subsection{Periodic \NHIMs}\label{subsec:periodicnhim}
\begin{figure}
    \includegraphics{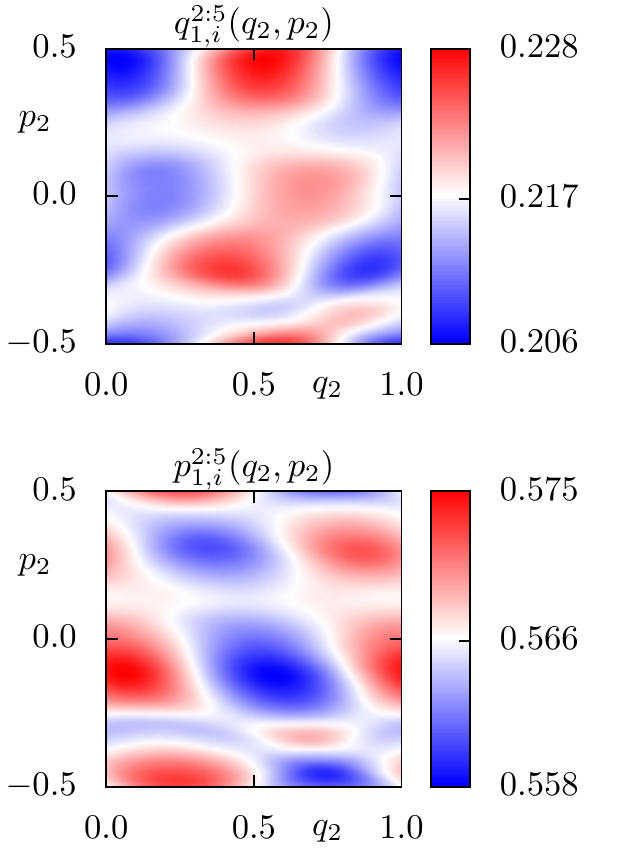}
    \caption{Parametrizations
        $q^{2:5}_{1,i}(q_2, p_2)$ and $p^{2:5}_{1,i}(q_2, p_2)$
        of a \twoD{} manifold, \prettyref{eq:nhimpara},
        of the periodic \NHIM{} \nhimpomnmaxi{2:5}{} for $\coup = 0.05$
        with $i$ corresponding to the periodic point with largest value of $p_1$.
        The corresponding periodic point
        of \mappingtwod{} is $(q, p) \approx (0.217, 0.566)$.
        The projection of these parametrizations
        onto the $(q_1, p_1)$ space
        are shown in the
        inset of \prettyref{fig:pbnhimprojpreimage}.
    }\label{fig:nhim_coord_surf}
\end{figure}

We introduce
periodic \NHIMs{} as the generalization of hyperbolic periodic orbits of the \twoD{} map.
If the coupling \coup{} is set to zero, one obtains
the Cartesian product of each hyperbolic periodic point
$\pomni{m:n}{i} = (q^{m:n}_i, p^{m:n}_i)$ of \mappingtwod{} in
$(q_1, p_1)$ space and the \twoD{} torus of the $(q_2, p_2)$ space.
This product structure is illustrated
in \prettyref{fig:ps_product}.
It gives rise to \twoD{} manifolds
$ \nhimpomn{m:n}{i} = \lbrace (q_{i}^{m:n},  p_{i}^{m:n}, q_2, p_2) \;|\; q_2 \in [0,1) , p_2 \in [-0.5,0.5) \rbrace$
in the \fourD{} phase space.
Each of the manifolds
$\nhimpomn{m:n}{i}$
is mapped to the next manifold,
$\mappingfourd(\nhimpomn{m:n}{i}) = \nhimpomn{m:n}{(i+1)\mod n}$,
and is thus invariant under the $n$-fold map,
$\mappingfourd^{n} (\nhimpomn{m:n}{i}) =  \nhimpomn{m:n}{i}$.
We denote this set of \twoD{} manifolds by
\begin{equation}
\nhimpomn{m:n}{} = \{ \nhimpomn{m:n}{i} \;|\; i = 0, \dots, n-1\}
,
\end{equation}
which is invariant
under one iteration of the map,
\begin{equation}
	\mappingfourd (\nhimpomn{m:n}{}) =  \nhimpomn{m:n}{}
	.
\end{equation}
It is normally hyperbolic due to the underlying
hyperbolic periodic orbit.
We call $\nhimpomn{m:n}{}$ a \emph{periodic NHIM},
as it is a \NHIM{} that is build from a periodic set of manifolds.

These periodic \NHIMs{} come in pairs, denoted by
\nhimpomnmax{m:n} and \nhimpomnmin{m:n},
as they emerge from the periodic orbits
\pomnmax{m:n} and \pomnmin{m:n}
of the map \mappingtwod.
It is required that the periodic orbit of \mappingtwod{} is hyperbolic or inverse hyperbolic.
For approximants of the golden torus this is the case for
$\kone > \ktwod_{c} \approx 0.97$ and sufficiently large period $n$.
Specifically, for $\kone=1.75$ this is the case for $n \ge 3$.
For an elliptic periodic orbit of \mappingtwod{}, one cannot construct a periodic \NHIM{}.

The case of small coupling,
$\coup \ll 1$,
can be considered as a small perturbation
of the case without coupling
such that the \NHIMs{} persist. This is the case as long as the hyperbolicity
in the normal directions is larger than in the tangential direction
\cite{Fen1972, HirPugShu1977, Man1978, Wig1994, Eld2013}.
Here it holds for ${\ktwo \ll \kone}$ and $\coup \ll 1$.
The periodic \NHIM{} for $\xi>0$ is no longer given by the trivial product structure.
Instead, we parametrize the \twoD{} manifolds by
\begin{equation}
\begin{aligned}
    \nhimpomn{m:n}{i} =
\lbrace & \bigl(q^{m:n}_{1, i}(q_2, p_2), p^{m:n}_{1, i}(q_2, p_2), q_2, p_2 \bigr)
\; | \;  \\
& q_2 \in [0, 1), \:  p_2 \in [-0.5, 0.5)
\rbrace ,
\end{aligned}
\label{eq:nhimpara}
\end{equation}
for $i=0,\dots,n-1$.
For $\coup \ll 1$
the parametrizations $q^{m:n}_{1, i}(q_2, p_2)$ and $p^{m:n}_{1, i}(q_2, p_2)$
are slightly varying around
the coordinates of the periodic point
$\pomni{m:n}{i}$
of \mappingtwod{}.
For an example this is visualized in \prettyref{fig:nhim_coord_surf}.
Numerically, we determine points on the periodic \NHIMs{}
by a binary contraction method \cite{BarJunFelMaiHer2018}
(for an overview of methods see Ref.~\cite{GonJun2022})
taking into account the periodic structure~\cite{Fir2020}.

Note that there is interesting dynamics on \NHIMs{}
\cite{LiShoTodKom2006a,GuzLegFro2009b,GonDroJun2014, GonJun2015, TerTodKom2015, CanHar2016, GonJun2022}.
This also holds for periodic \NHIMs{}
and is presented for increasing periods
in \prettyref{app:dynonnhim}.

\subsection{Partial barriers based on periodic \NHIMs}\label{subsec:flux4d}
We consider \threeD{} partial barriers $C$ connecting the \twoD{} manifolds
of the periodic \NHIMs{} \nhimpomnmin{m:n} and \nhimpomnmax{m:n}
in an alternating manner,
generalizing from the \twoD{} map to the \fourD{} map.
Without coupling ($\coup=0$) a partial barrier is formed by the
Cartesian product of
a \oneD{} partial barrier of the \twoD{} map
and the \twoD{} torus of the $(q_2, p_2)$ space.
It constitutes a \threeD{} manifold dividing
the \fourD{} phase space.
This product structure is illustrated
in \prettyref{fig:ps_product}.
The partial barrier and its preimage enclose $2n$
\fourD{} lobe volumes,
which are a Cartesian product of \twoD{} lobes
in the $(q_1, p_1)$ space and the
\twoD{} torus of the $(q_2, p_2)$ space,
see \prettyref{fig:ps_product}.

In the presence of a small coupling,
$\coup \ll 1$,
the \NHIMs{} persist and a slightly varied partial barrier $C$
can be constructed by connecting the \NHIMs{} (in infinitely many ways).
One has to ensure that
the partial barrier and its preimage
intersect in the periodic \NHIMs{} only,
i.e.\ $C \cap \mappingfourd^{-1} (C) = \nhimpomnmin{m:n}{} \cup \nhimpomnmax{m:n}$.

In order to determine the flux across partial barriers of this kind
we apply the flux formula, \prettyref{eq:flux_general}.
Since the \fourD{} standard map~\mappingfourd{} is exact volume preserving with
the three-form
\begin{equation}
	\minusoneform =  p_1 \; \ud q_1 \wedge \ud p_2 \wedge \ud q_2 .
	\label{eq:threeform}
\end{equation}
a generating two-form $\gentwoformfourd$ exists,
given by
\begin{equation}
	\gentwoformfourd
	=
	\Big(\frac12 p_1^2 - V(q_1 + p_1, q_2 + p_2)  \Big) \; \ud p_2 \wedge  \ud q_2
	,
\label{eq:gentwostd4d}
\end{equation}
and fulfilling \prettyref{eq:volprev}.
The invariant set $\invariantset{}$ in
the flux formula,
\prettyref{eq:flux_general},
consists of $\nhimpomnmax{m:n}$ and $\nhimpomnmin{m:n}$
at the ends of $U$,
contributing with different signs,
$\invariantset{} = \nhimpomnmax{m:n} - \nhimpomnmin{m:n}$.
This leads for the \fourD{} case to the
explicit expression
\begin{equation}
\flux^{\fourD{}}
= \int \limits_{B} \gentwoformfourd{}
=
\sum_{i = 0}^{n-1}
\int \limits_{\nhimpomnmaxi{m:n}{i}}
\gentwoformfourd
\; - \;
\sum_{i = 0}^{n-1}
\int \limits_{\nhimpomnmini{m:n}{i}}
\gentwoformfourd
.
\label{eq:fluxfourd}
\end{equation}
Let us stress, that
the \fourD{} flux across such a partial barrier
is entirely determined by integrating a generating two-form \gentwoform{}
over the \twoD{} manifolds of the periodic \NHIMs{}.
Thus the flux does not depend on the specific choice of partial barrier.

For the chosen three-form \minusoneform{}, \prettyref{eq:threeform},
and the related generating two-form  \gentwoformfourd{}, \prettyref{eq:gentwostd4d},
just integrals with respect to the \twoD{} torus of the $(q_2, p_2)$ space
have to be determined.
Numerically, one can use
a grid in $(q_2, p_2)$ space and
the parametrization of \nhimpomn{m:n}{} from \prettyref{eq:nhimpara}.
The flux~$\flux^{\fourD}$ is determined for $\coup=0.05$
and various periods, see
\prettyref{fig:flux_4d_periods}
and \prettyref{tab:flux_4d}.

Note that without coupling, $\coup=0$,
the evaluation of the flux formula, \prettyref{eq:fluxfourd},
simplifies, since the generating two-form \gentwoformfourd{}, \prettyref{eq:gentwostd4d},
and the \NHIMs{}
become independent of $q_2$ and $p_2$.
Consequently, the integration with respect to these coordinates simply yields $1$,
i.e.\ the area of the \twoD{} torus of the $(q_2, p_2)$ space.
The remaining sum is equivalent to the flux formula for \twoD{} maps, \prettyref{eq:flux_twod}.

\subsection{Cantorus-NHIM}\label{sec:cantorus_nhim}

We now establish for \fourD{} maps what we call a \emph{cantorus-NHIM}.
Without coupling, $\coup=0$,
it is a Cartesian product of a cantorus with irrational frequency $\nu$ of the \twoD{} map in $(q_1, p_1)$ space
and the \twoD{} torus of the $(q_2, p_2)$ space.
It is normally hyperbolic due to the properties of the cantorus
and it is invariant.
We expect that this \NHIM{}
with a Cantor set structure
persists for small coupling, $\coup \ll 1$,
just like other \NHIMs{}.
It can be successively approximated by the pair of periodic \NHIMs{},
\nhimpomnmin{m:n} and \nhimpomnmax{m:n},
with frequencies $\frac{m}{n}$ obtained by successive
truncations of the continued fraction expansion of the irrational
frequency $\nu$.

\begin{figure}
    \includegraphics{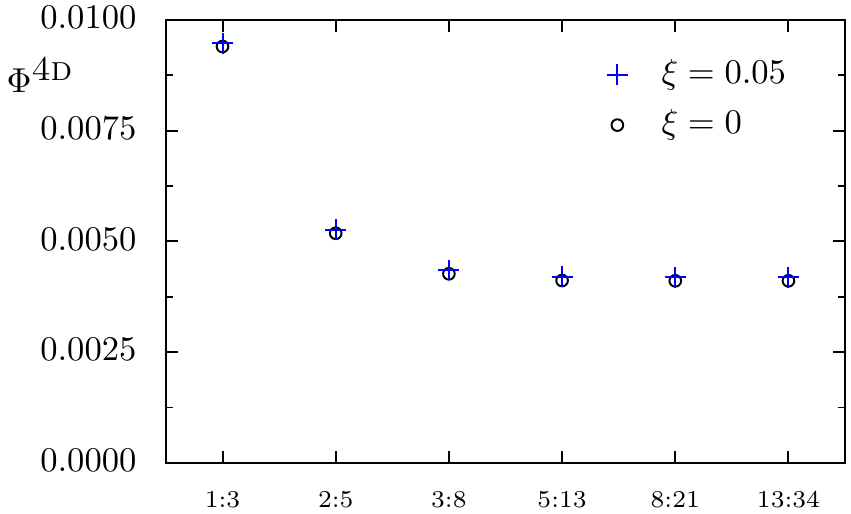}
    \caption{
        Flux $\flux^{\fourD}$, \prettyref{eq:fluxfourd},
        across partial barriers based on periodic \NHIMs{}
        \nhimpomn{1:3}{} to \nhimpomn{13:34}{}
        for $\coup = 0.05$ (crosses),
        converging to the flux across the cantorus-\NHIM{}.
        It is compared to the
        flux $\flux^{\fourD}(\coup = 0)$ without coupling (circles)
        corresponding to $\flux^{\twoD}$.
    }\label{fig:flux_4d_periods}
\end{figure}

The flux~$\flux^{\fourD}$ across a partial barrier based on a
cantorus-\NHIM{} is, in analogy to \twoD{} maps,
approximated by the flux
$\flux^{\fourD}$, \prettyref{eq:fluxfourd},
across partial barriers based on
the pair of periodic \NHIMs{}
\nhimpomnmin{m:n} and \nhimpomnmax{m:n}.
Fast convergence of the flux~$\flux^{\fourD}$ with increasing periods $n$
is shown in \prettyref{fig:flux_4d_periods} and
\prettyref{tab:flux_4d}
for \mappingfourd{} with $\coup=0.05$.
The flux shows the same qualitative dependence on the period $n$
and is slightly larger than the flux
without coupling
$\flux^{\fourD}(\coup=0)$,
which corresponds to
$\flux^{\twoD}$
for \mappingtwod.

\begin{table}
    \begin{tabular}{ c| c | c}
    $m\!:\!n$ & $\flux^{\fourD}(\coup = 0)$ &$\flux^{\fourD}(\coup = 0.05)$   \\
    \hline
    $1\!:\!3$ &   $0.0094004$ & $0.0094771$ \\
    $2\!:\!5$ &   $0.0051837$ & $0.0052659$ \\
    $3\!:\!8$ &   $0.0042684$ & $0.0043532$ \\
    $5\!:\!13$ &  $0.0041162$ & $0.0042000$ \\
    $8\!:\!21$ &  $0.0041104$ & $0.0041941$ \\
    $13\!:\!34$ & $0.0041104$ & $0.0041941$ \\
\end{tabular}
    \caption{Flux $\flux^{\fourD}$,
    	as displayed in
    	\prettyref{fig:flux_4d_periods},
    	showing fast convergence with increasing period $n$.
    }
    \label{tab:flux_4d}
\end{table}

We expect that these partial barriers based on a cantorus-\NHIM{}
are as important in \fourD{} maps as their counterparts in \twoD{} maps.
This generalization to \fourD{} maps should also apply
to other cantori of the \twoD{} map,
e.g., originating from rotational \oneD{} tori with
different irrational frequency $\nu$ or from oscillatory \oneD{} tori.

\begin{figure}
    \includegraphics{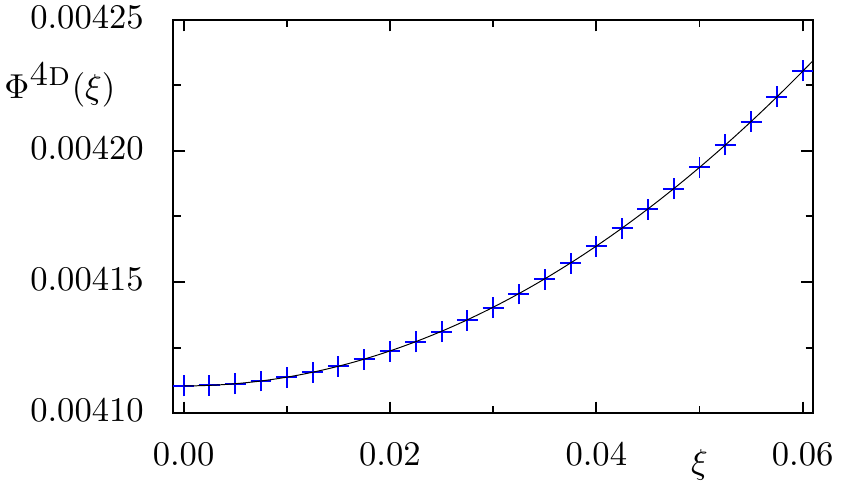}
    \caption{
        Flux $\flux^{\fourD}$, \prettyref{eq:fluxfourd},
        across partial barriers based on periodic \NHIMs{}
        \nhimpomn{8:21}{}
        vs.\ coupling $\coup$ (crosses),
        showing a quadratic increase (solid line).
    }
    \label{fig:flux_4d_coupling}
\end{figure}
We investigate the dependence of the flux $\flux^{\fourD}$
on the coupling $\coup$ for the example
of partial barriers based on the periodic \NHIM{} \nhimpomn{8:21}{},
which closely approximates the flux across the cantorus-\NHIM{}.
In \prettyref{fig:flux_4d_coupling} the flux $\flux^{\fourD}$
shows for $\coup \in [0.0, 0.06]$
a quadratic increase,
$\flux^{\fourD} (\coup) \approx \flux^{\fourD}(\coup = 0) + 0.0333 \cdot \coup^2$.
For other parameters \kone{} and different periods we find
also a quadratic dependence (not shown).

\subsection{4D lobe volume}\label{subsec:lobe4d}

\begin{figure*}
	\includegraphics{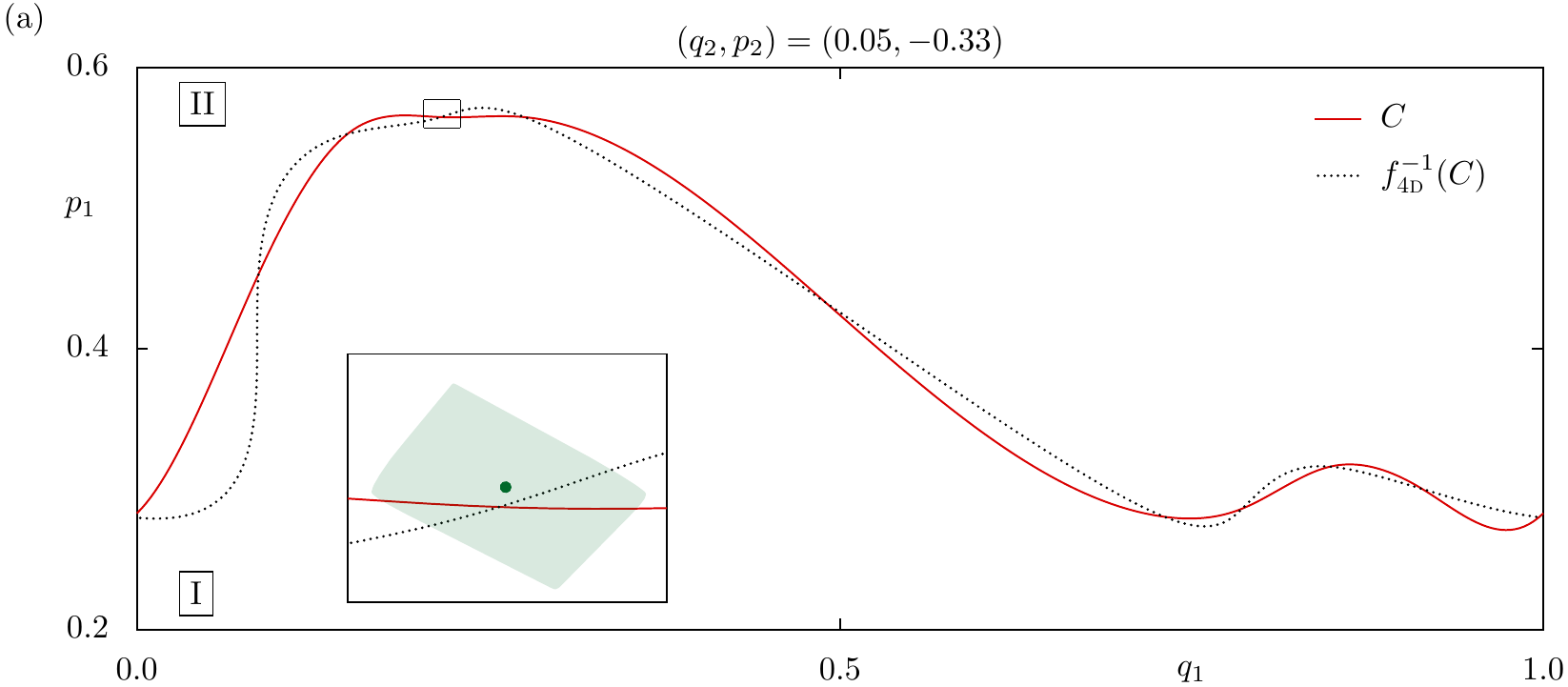}

        \vspace*{3ex}
	\includegraphics{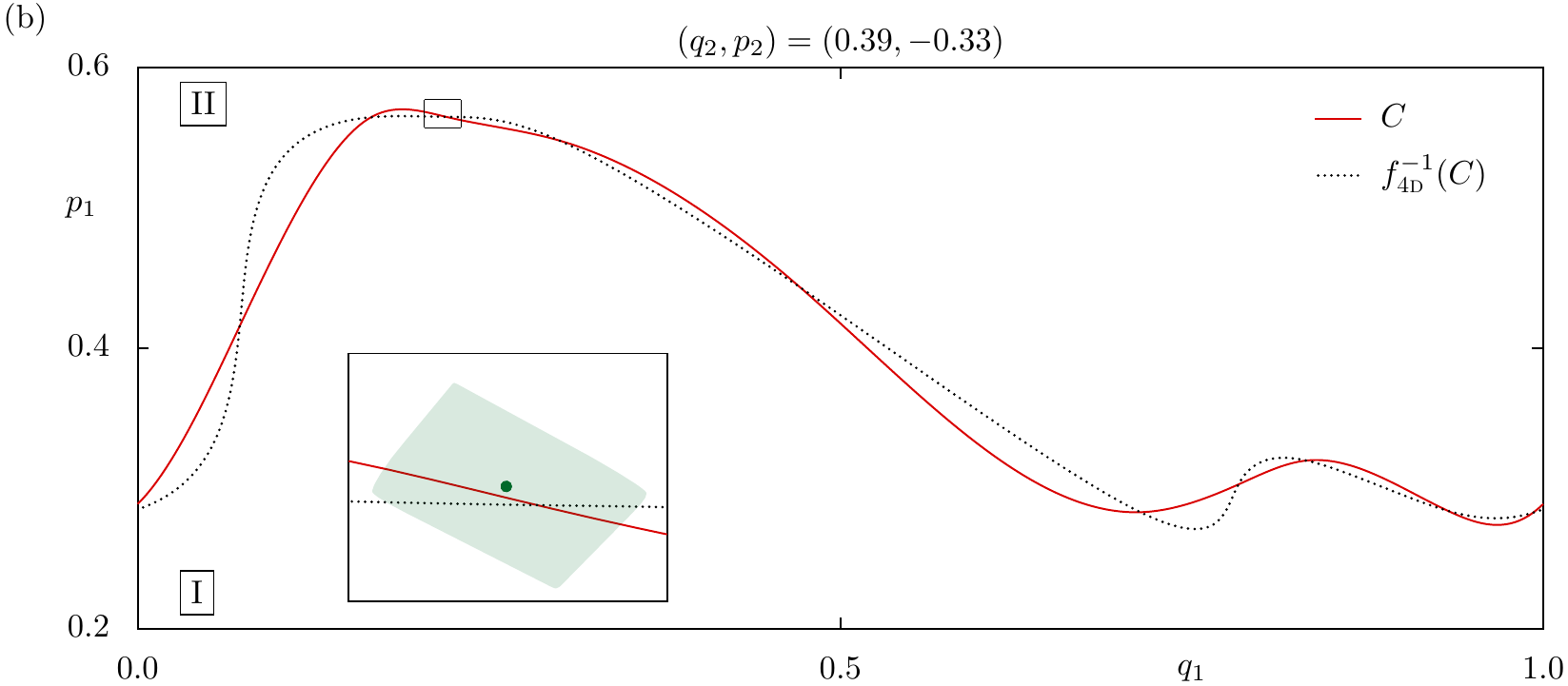}
	\caption{
		Example of a \threeD{} partial barrier $C$ (red solid line)
		and its preimage $\mappingfourd^{-1}(C)$ (black dotted line)
		in a $(q_2, p_2)$ section of phase space
		for (a) $(q_2, p_2) = (0.05, -0.33)$
		and (b)  $(q_2, p_2) = (0.39, -0.33)$,
		intersecting at periodic \NHIMs{} \nhimpomnmin{2:5} and \nhimpomnmax{2:5}
		for $\coup = 0.05$
		and confining $5$ lobes each in region~\regionone{}
		and~\regiontwo{}.
		Insets: Magnification of one intersection
		together with
		the projection of \nhimpomnmax{2:5}
		on the $(q_1, p_1)$ space (green colored area),
		see also \prettyref{fig:nhim_coord_surf},
		and the corresponding periodic point $\pomnmax{2:5}$
		of the \twoD{} map (green point).
	}
	\label{fig:pbnhimprojpreimage}
\end{figure*}

\begin{figure*}
	\includegraphics{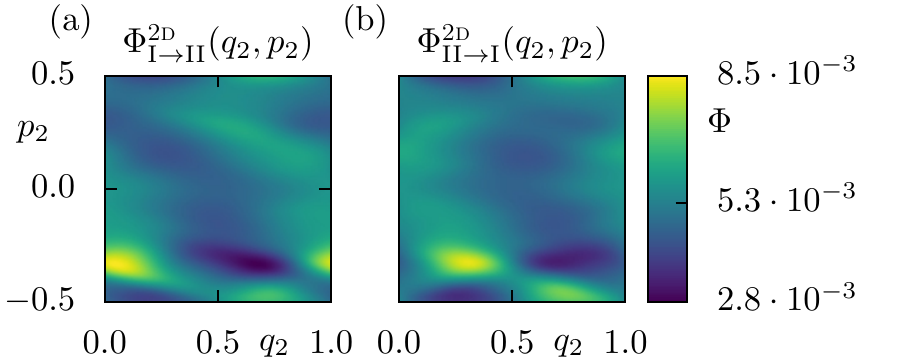}
	\caption{
		Flux
		(a) $\flux^{\textsc{2d}}_{\regionone \rightarrow \regiontwo}$
		and
		(b) $\flux^{\textsc{2d}}_{\regiontwo \rightarrow \regionone}$
		depending on the $(q_2, p_2)$ section
		for a \threeD{} partial barrier based
		on the periodic \NHIMs{} \nhimpomn{2:5}{}
		for $\coup = 0.05$.
	}\label{fig:flux_plane_both}
\end{figure*}

Complementary to the evaluation of the flux formula,
\prettyref{eq:fluxfourd},
we want to explicitly determine the \fourD{} lobe volume
for an example of a \threeD{} partial barrier based on a periodic \NHIM{}.
This has several motivations:
(i) it gives an independent verification of \prettyref{eq:fluxfourd},
(ii) it provides an impression of the geometry of the lobes,
as for $\coup > 0$
one no longer has a product structure,
and (iii) it is a preparation for the determination of a local flux introduced and studied
in \prettyref{sec:localflux}.

For convenience
we consider \threeD{} partial barriers that can be written
as a function $p_1^\text{pb} (q_1, q_2, p_2)$
of the three cyclic variables $(q_1, q_2, p_2)$ of \mappingfourd{}.
This function can be expressed by a finite Fourier series.
See Ref.~\cite{Fir2020} for a construction, where
a specific partial barrier is chosen by minimizing
its curvature along the $q_1$ direction.
One has to check numerically that
the partial barrier and its preimage intersect
in the periodic \NHIMs{} only, i.e.\
$C \cap \mappingfourd^{-1} (C) = \nhimpomnmin{m:n}{} \cup \nhimpomnmax{m:n}$.

As an example we choose a \threeD{} partial barrier based on the
periodic \NHIMs{} \nhimpomnmin{2:5} and \nhimpomnmax{2:5}.
Sections for two exemplary fixed values of $(q_2, p_2)$ of the partial barrier
and its preimage are shown in
\prettyref{fig:pbnhimprojpreimage}.
In such a $(q_2, p_2)$ section
the dimensionality of the objects is reduced by two,
i.e.\
the partial barrier and its preimage appear as \oneD{} lines.
They confine \twoD{} lobe areas, namely
$5$ lobes in region~\regionone{}
and $5$ lobes in region~\regiontwo{}
with the same topology as in the uncoupled case, compare
\prettyref{fig:ps_product}.
The area of the individual lobes depends on the $(q_2, p_2)$ section,
e.g., the area of the left-most lobe is bigger in
\prettyref{fig:pbnhimprojpreimage}(a)
than in
\prettyref{fig:pbnhimprojpreimage}(b).

The sum of the lobe areas
in region~\regionone{}
in a given $(q_2, p_2)$ section
gives the \twoD{} flux  to region~\regiontwo{},
denoted by
$\flux^\twoD_{\regionone \rightarrow \regiontwo}(q_2, p_2)$.
The corresponding flux in the reverse direction is denoted by
$\flux^\twoD_{\regiontwo \rightarrow \regionone}(q_2, p_2)$.
They are calculated according to
\prettyref{eq:lobe_area}
in \prettyref{app:lobearea}.
Note, that
under one iteration of the map \mappingfourd{}
these lobe areas are no longer contained in the same
$(q_2, p_2)$ section.

There is an asymmetry of this \twoD{} flux
for a given $(q_2, p_2)$ section.
In \prettyref{fig:pbnhimprojpreimage}(a)
one can observe that the flux
$\flux^\twoD_{\regionone \rightarrow \regiontwo}(q_2, p_2)$
is greater than $\flux^\twoD_{\regiontwo \rightarrow \regionone}(q_2, p_2)$,
whereas
in \prettyref{fig:pbnhimprojpreimage}(b)
it is the other way around.
This is illustrated quantitatively in \prettyref{fig:flux_plane_both}
for all sections $(q_2, p_2)$.
Thus, one concludes that zero net flux does not hold for
each $(q_2, p_2)$ section individually.

Finally, the \fourD{} flux across the \threeD{} partial barrier
is obtained by integrating the \twoD{} flux over all sections,
\begin{equation}
	\flux^\fourD =  \iint \flux^\twoD{}(q_2, p_2) \: \ud q_2 \: \ud p_2 ,
	\label{eq:flux4dint}
\end{equation}
where it should be irrelevant whether
$\flux^\twoD_{\regionone \rightarrow \regiontwo}(q_2, p_2)$
or $\flux^\twoD_{\regiontwo \rightarrow \regionone}(q_2, p_2)$
is integrated, as for the flux~$\flux^\fourD$
we have zero net flux, see \prettyref{sec:pbfluxgen}.

For a \threeD{} partial barrier based on the
periodic \NHIMs{} \nhimpomnmin{2:5} and \nhimpomnmax{2:5}
with $\coup=0.05$
we obtain,
$\flux^{\fourD}_{\regionone \rightarrow \regiontwo} =
\flux^{\fourD}_{\regiontwo \rightarrow \regionone} = 0.0052659$.
This result of the volume measurement of the \fourD{} lobe volumes
confirms the result
$\flux^{\fourD} = 0.0052659$
of the flux formula, \prettyref{eq:fluxfourd},
see \prettyref{tab:flux_4d},
entirely determined
by the periodic \NHIMs{},
within the numerical accuracy of the determination of the \NHIMs{}.

\subsection{Partial barrier between resonance channels}\label{subsec:resonance_channel}

We now discuss where the partial barriers studied in this paper
are located in phase space and frequency space,
in particular in comparison to resonance channels.
This will be helpful
for understanding more generic partial barriers,
see the discussion in the outlook, \prettyref{sec:outlook}.
Quite importantly, it explains
why the flux~$\flux^{\fourD}$ studied so far,
will be considered as a global flux,
in contrast to a local flux introduced in the next
\prettyref{sec:localflux}.

We start with the important concept of resonances (for the \twoD{} standard map)
and resonance channels (for the \fourD{} map).
Resonances of a \twoD{} map correspond to rational frequencies.
For the \twoD{} standard map at $K=0$ lines with rational $p$
break-up for $K>0$ into a chain of elliptic and hyperbolic
periodic points together with surrounding regular tori and
chaotic layer, respectively, to form a so-called resonance.
The two most prominent resonances have
frequency $\nu = 0$ at $p=0$
and $\nu = \frac{1}{2}$ at $p=0.5$,
see \prettyref{fig:ps2d}.
The corresponding elliptic points are surrounded by a regular island
and the hyperbolic points are embedded in a chaotic layer.
Transport between these chaotic layers is possible
for $K > \kcrit$. It is limited by
the partial barrier with the smallest flux
between the resonances
$\nu=0$ and $\nu=\frac{1}{2}$
which is based on the golden cantorus
with $\nu \approx 0.382$,
see \prettyref{sec:pbflux2d}.

For a \fourD{} map the \twoD{} regular tori are characterized
by two frequencies $(\nu_1, \nu_2)$.
A pair of frequencies is called resonant if it fulfills
$m_1 \nu_1 + m_2 \nu_2 = k$ with integers $m_1, m_2, k$.
In frequency space they form so-called resonance lines, denoted by $m_1:m_2:k$.
In phase space the corresponding objects constitute a resonance channel
which consists of families of \twoD{} tori, families of elliptic \oneD{} tori
and hyperbolic \oneD{} tori, together with their surrounding
chaotic motion, see\cite{OnkLanKetBae2016}
for an illustration of a generic resonance channel of a \fourD{} map.

For the \fourD{} map with $\kone>0$ and $K_2=0$ and small coupling
the resonances of \mappingtwod{}
for $\nu=0$ and $\nu=\frac{1}{2}$ lead to the $1:0:0$ resonance line
and the $2:0:1$ resonance line,
with frequencies $\nu_1=0$ and $\nu_1=\frac{1}{2}$,
respectively,
while $\nu_2 \in [-0.5, 0.5)$ extends along the entire $\nu_2$ axis.
These resonance lines are thus both parallel to the $\nu_2$ axis
in frequency space.
In phase space, both resonance lines correspond to resonance channels
extending in $p_2$ direction.
For $\kone > \kcrit$ transport across a partial barrier between these resonances
is possible for \mappingtwod{} and therefore also
across a partial barrier between the corresponding resonance channels
for \mappingfourd{}.
The flux between these resonance channels
is restricted by the cantorus-\NHIM{},
as studied in \prettyref{sec:flux4d}.
It gives rise to a partial barrier extended between the
${1:0:0}$ and ${2:0:1}$ resonance lines in frequency space,
such that $\nu_1 = 1-\sigma_{\text{G}}$
and $\nu_2$ is extended along the entire $\nu_2$-axis.
This corresponds in phase space to
the entire $p_2$-axis with $p_2 \in [-0.5, 0.5)$.
The flux~$\flux^\fourD$ across the cantorus-\NHIM{}
from resonance channel
${1:0:0}$ to resonance channel ${2:0:1}$
is thus a \emph{global flux}.

Note that partial barriers associated with irrational frequency lines
have been proposed in Ref.~\cite{MarDavEzr1987}.
To our knowledge, the partial barrier based on the cantorus-\NHIM{}
is the first explicit realization.

\section{Local Flux in \fourD{} maps}\label{sec:localflux}
\subsection{Motivation}

Resonance channels, introduced in
\prettyref{subsec:resonance_channel},
play a prominent role
for chaotic transport in \fourD{} symplectic maps.
Transport along a resonance channel is often slow due
to Arnold diffusion~\cite{Arn1964, Chi1979, Loc1999, Dum2014}.
Transport from one resonance channel to another may occur
at so-called resonance junctions\cite{Hal1995, Hal1999, GotNoz1999, Got2006, EftHar2013, KarKes2018}.
Here we are interested in
transport between (nearly) parallel resonance channels
restricted by a partial barrier in between the resonance channels.
Due to the slow transport along a resonance channel,
the local properties of the partial barrier
at a given position along the resonance channel
are important.
Typically, these local properties will be more relevant than the global flux
accumulated along the entire resonance channel.
We will therefore introduce a \emph{local flux},
that depends on the position along
a resonance channel.

Specifically, for the \fourD{} standard map \mappingfourd{}
with $K_2=0$ and small coupling $\coup \ll 1$,
transport in the $p_2$-direction
is rather slow. This is the case as the change in $p_2$
is proportional to the coupling~\coup{},
see Eqs.~(\ref{eq:map_stdfourd}) and (\ref{eq:potential_stdfourd}).
The dynamics in $q_1$ and $q_2$ is much faster.
Also transitions in $p_1$ from the
${1:0:0}$ resonance channel
to the ${2:0:1}$ resonance channel
via a partial barrier
may occur many times,
while the $p_2$ coordinate barely changes.

Thus, in order to describe chaotic transport more precisely,
one needs to know the local flux
near a given $p_2$ coordinate along the resonance channel,
rather than the global flux~$\flux^\fourD$.
To this end we
use a $p_2$-section of the \fourD{} lobe volume to
define the local \threeD{} flux~$\flux^\threeD (p_2)$
across a partial barrier.
It can be determined based on the
area of the lobe in $(q_2, p_2)$ sections,
$\flux^\twoD_{\regionone \rightarrow \regiontwo}(q_2, p_2)$
or $\flux^\twoD_{\regiontwo \rightarrow \regionone}(q_2, p_2)$,
as introduced in \prettyref{subsec:lobe4d}.
By integrating over $q_2$ one finds the local flux,
\begin{equation}
\flux^\threeD (p_2) = \int  \ud q_2 \: \flux^\twoD(q_2, p_2),
\label{eq:fluxthreed}
\end{equation}
which differs from \prettyref{eq:flux4dint}
as there is no integration along $p_2$.
As an aside, we mention that
the local flux~$\flux^\threeD (p_2)$
divided by the \threeD{} volume of the
resonance channel at the given $p_2$ coordinate
gives an estimate for the local transition rate
across the partial barrier.

Note that for \mappingfourd{} without coupling, $\coup = 0$,
the flux~$\flux^\twoD(q_2, p_2)$
does not depend on $q_2$ and $p_2$,
due to the product structure,
see \prettyref{fig:ps_product}.
Therefore the integration over
$q_2 \in [0, 1)$ in \prettyref{eq:fluxthreed}
simply gives a factor of $1$
and the local flux $\flux^\threeD$ does not depend on $p_2$
in this case,
leading to the same values for $\flux^\threeD(\coup = 0)$
and $\flux^\fourD(\coup = 0)$.

\begin{figure}[h!]
	\includegraphics{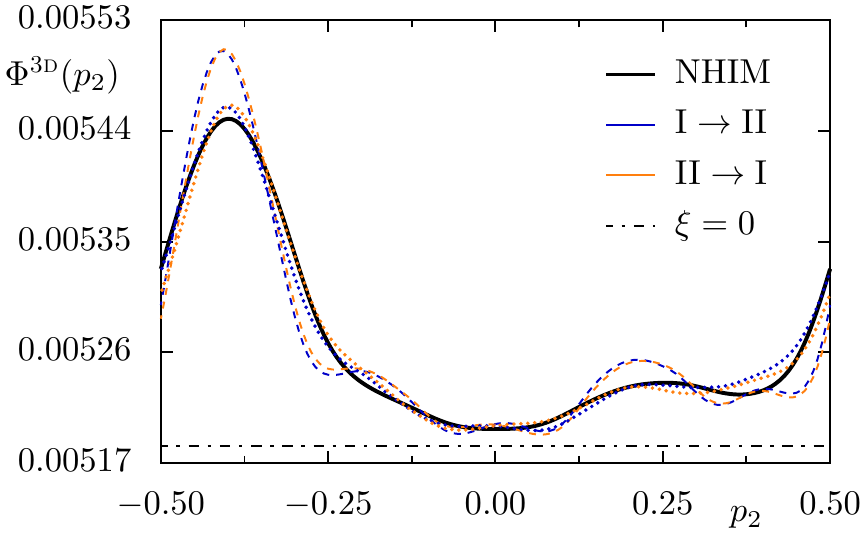}
	\caption{
		Local flux $\flux^\textsc{3d} (p_2)$, \prettyref{eq:fluxthreed},
		for two different
		\threeD{} partial barriers
		(dotted and dashed thin lines)
		and both directions $\regionone \rightarrow \regiontwo$
		(dark blue)
		and $\regiontwo \rightarrow \regionone$
		(light orange)
		based on the peridic \NHIM{} \nhimpomn{2:5}{}
		for coupling $\coup = 0.05$.
		Comparison to the local
		flux without coupling, $\flux^{\textsc{3d}}(\coup = 0)$
		(black dash-dotted line)
		and to the \NHIM{} contribution,
		$\flux^\threeD_\text{NHIM}(p_2)$,
		\prettyref{eq:nhimcont}
		(black thick line).
	}
	\label{fig:flux_p2_compare}
\end{figure}

\subsection{Dependence on partial barrier}\label{subsec:localfluxpb}

For a given periodic \NHIM{} or cantorus-\NHIM{}
the global flux $\flux^\fourD$ is independent of a specific partial barrier,
see \prettyref{sec:flux4d}.
Moreover, the net flux is zero across any partial barrier
for the considered exact volume preserving maps.
Here, we investigate whether the
local flux~$\flux^\threeD (p_2)$, \prettyref{eq:fluxthreed},
depends on specific partial barriers and
examine whether zero net flux holds locally.
Furthermore, we want to study how the local flux depends on $p_2$.

To this end we consider two (arbitrary) examples of partial barriers.
The first was already utilized in  Sec.~\ref{subsec:lobe4d}.
The second partial barrier is defined by
requiring a specific slope of the partial barrier at the periodic
\NHIMs{} in addition to minimizing its curvature along the
$q_1$ direction~\cite{Fir2020}.
The lobe areas in the  $(q_2, p_2)$ sections differ substantially
and one obtains a different flux $\flux^\textsc{2d}(q_2, p_2)$
than in \prettyref{fig:flux_plane_both} (not shown).
We checked that the global flux $\flux^\fourD$ is indeed
the same as before.

In \prettyref{fig:flux_p2_compare} we show
the local flux $\flux^\threeD (p_2)$
from region~$\regionone{}$ to region~$\regiontwo{}$ and vice versa
for both partial barriers.
This allows to draw several conclusions on
the local flux $\flux^\threeD (p_2)$:
(i) it depends on $p_2$,
(ii) it depends on the partial barrier,
and
(iii) it weakly depends on the direction
($\regionone{} \to \regiontwo{}$ vs.\ $\regiontwo{} \to \regionone{}$),
i.e.\ there is no zero net flux
in a given $p_2$ section.

Quite importantly, \prettyref{fig:flux_p2_compare} exhibits also common features
of the local flux $\flux^\threeD (p_2)$.
For both partial barriers and both directions
we find a prominent maximum at $p_2 \approx -0.4$
and a minimum at $p_2 \approx 0$
(with larger values than
without coupling, $\coup = 0$,
for all $p_2$).
This will be further studied below.

Note that the definition of a local
flux $\flux^\threeD (p_2)$ can be done in various ways.
In \prettyref{eq:fluxthreed} the $p_2$ section of exit lobes is used.
Alternatively, one could take the $p_2$ section of the iterated exit lobes.
In this case we find qualitatively similar, but quantitatively different results.

\subsection{NHIM contribution}
The local flux, although depending on specific partial barriers,
shares common features, as shown in \prettyref{fig:flux_p2_compare}.
Common to both partial barriers is, that they are based on
the same periodic \NHIM{}.
Thus, it is plausible that these common features
are related to the underlying periodic \NHIM{}.

We propose a splitting of the local flux $\flux^\threeD (p_2)$
into two contributions,
\begin{equation}
\flux^\threeD (p_2) = \flux^\threeD_\text{NHIM} (p_2) + \delta_\text{pb} (p_2)
,
\label{eq:flux_threed_splitting}
\end{equation}
where the first term $\flux^\threeD_\text{NHIM} (p_2)$,
defined
in \prettyref{eq:nhimcont}
in \prettyref{app:nhimcont},
is entirely based on the periodic \NHIM{}.
We thus call it the \NHIM{} contribution.
The remaining term,
$\delta_\text{pb}(p_2) := \flux^\threeD (p_2) - \flux^\threeD_\text{NHIM} (p_2)$,
describes the deviations from the \NHIM{} contribution,
i.e.\ the additional $p_2$ dependence
specific for the considered partial barrier.

The \NHIM{} contribution,
$\flux^\threeD_\text{NHIM} (p_2)$,
is shown in
\prettyref{fig:flux_p2_compare}
and it
captures the common features of the $p_2$ dependence
of the local flux $\flux^\threeD (p_2)$ quite well.
We find that the local flux of partial barriers
based on \NHIMs{} with longer periods and varying parameters $(\kone, \coup)$,
are also well captured by the \NHIM{} contribution
(not shown).
For a different class of partial barrier
we also find good agreement, as discussed below
in Sec.~\prettyref{subsec:local_flux_mani}.
We thus conjecture,
that the $p_2$ dependence of the
local flux $\flux^\threeD (p_2)$,
\prettyref{eq:fluxthreed},
is in general well described by the \NHIM{} contribution,
$\flux^\threeD_\text{NHIM} (p_2)$,
\prettyref{eq:nhimcont}.

Numerically, we find
that integrating over $p_2$ of the \NHIM{} contribution
yields the global flux $\flux^\fourD{}$,
$\int \ud p_2 \:  \flux^\threeD_\text{NHIM} (p_2) = \flux^\fourD{}$,
as well as cancellation of the barrier-specific contribution,
$\int \ud p_2 \: \delta_\text{pb}(p_2) = 0$.
Both results are found up to the numerical precision of the \NHIMs{}
and we expect them to hold exactly.

\subsection{Cantorus-NHIM}
\begin{figure}
    \includegraphics{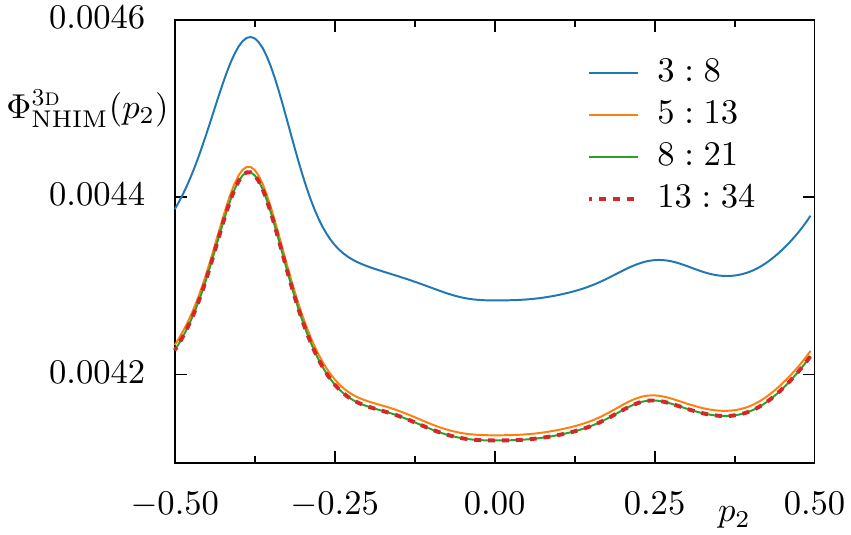}
    \caption{
        \NHIM{} contribution $\flux^\threeD_\NHIM (p_2)$,
        \prettyref{eq:nhimcont},
        to the local flux
        for periodic \NHIMs{}
        \nhimpomn{3:8}{} to \nhimpomn{13:34}{} (top to bottom)
        for coupling $\coup = 0.05$.
    }
    \label{fig:plot_flux_p2_increasing_period}
\end{figure}
\begin{figure}[h]
    \centering
    \includegraphics{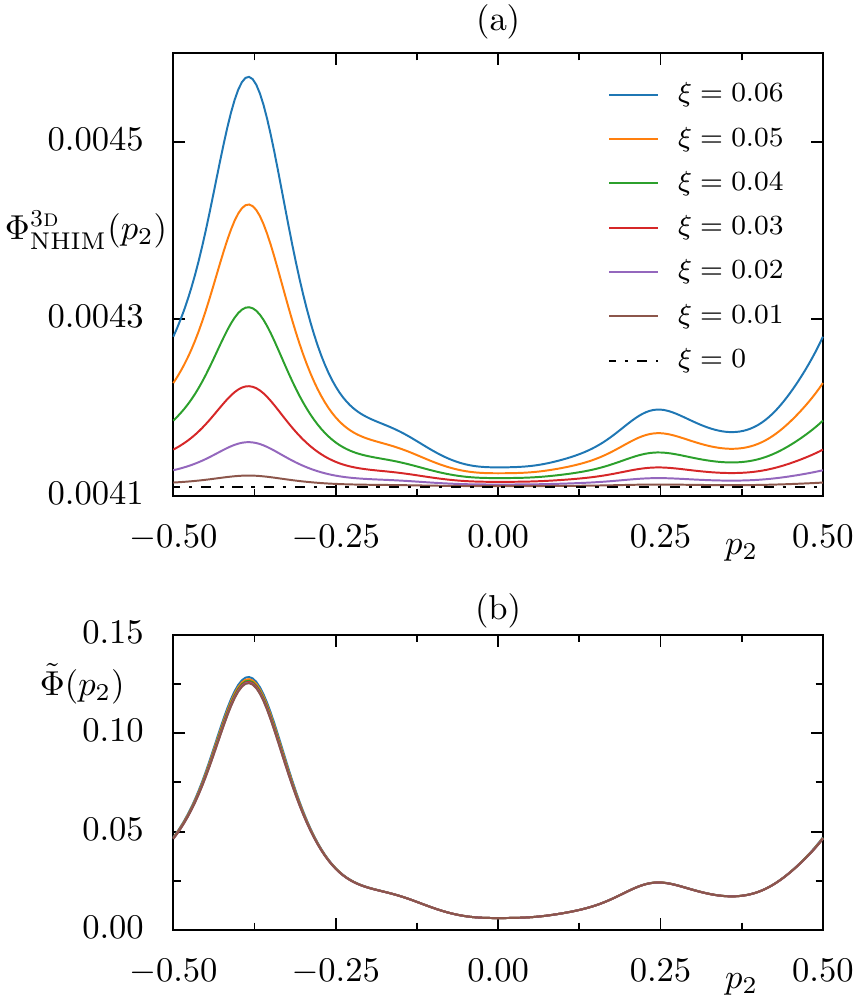}
    \caption{
    	(a)
    	Dependence of the
        \NHIM{} contribution $\flux^\threeD_\NHIM (p_2)$,
        \prettyref{eq:nhimcont},
        on the coupling $\coup$
        for the periodic \NHIM{}
        \nhimpomn{8:21}{} for $\coup \in \{0.01, \cdots, 0.06\}$
        (solid lines from bottom to top)
        and flux $\flux^{\threeD}(\coup = 0)$ without coupling
        (dash-dotted line).
        (b) Quadratically scaled flux
        $\tilde{\flux}(p_2) = [\flux^\threeD_\NHIM{} - \flux^{\threeD}(\coup = 0)] / \coup^2$.
    }
    \label{fig:plot_flux_p2_increasing_coupling}
\end{figure}

For a partial barrier based on a cantorus-\NHIM{}
the local flux~$\flux^{\threeD}(p_2)$
is approximated by the
\NHIM{} contribution, $\flux^\threeD_\text{NHIM} (p_2)$,
across partial barriers based on
the pair of periodic \NHIMs{},
\nhimpomnmin{m:n} and \nhimpomnmax{m:n},
with frequencies $\frac{m}{n}$
converging to the frequency of the cantorus-\NHIM{}.
We are interested in the convergence of the $p_2$ dependence,
which is presented
in \prettyref{fig:plot_flux_p2_increasing_period}.
Fast convergence of
$\flux^\threeD_\NHIM (p_2)$
with increasing periods $n$
can be observed.
As for $\frac{2}{5}$ in
\prettyref{fig:flux_p2_compare},
we find for other frequencies~$\frac{m}{n}$ a prominent peak located
at $p_2 \approx - \frac{m}{n}$,
converging towards $p_2 = \sigma_{\text{G}} - 1$,
and a minimum at $p_2 \approx 0$.
We observe that the prominent peak occurs at the $p_2$
value, where the dynamics on the \NHIM{} shows
the biggest resonance,
see \prettyref{fig:nhim_poincare_increasing_period}
in \prettyref{app:dynonnhim}.
This is at a crossing of the $1:1:0$ resonance channel
with the cantorus-\NHIM{}.

In \prettyref{sec:cantorus_nhim} a quadratic dependence of the
global flux $\flux^{\fourD}$ on the coupling $\coup$ is observed.
Here we study this dependence for the \NHIM{} contribution
$\flux^\threeD_\NHIM (p_2)$ of the local flux,
again for the periodic \NHIM{} \nhimpomn{8:21}{},
which closely approximates the flux across the cantorus-\NHIM{}.

In \prettyref{fig:plot_flux_p2_increasing_coupling}(a)
we observe qualitatively that
for all couplings $\coup$ the \NHIM{} contribution
has roughly the same $p_2$ dependence,
but with increasing amplitude and increasing average.
We quadratically scale the difference of the
\NHIM{} contribution $\flux^\threeD_\NHIM (p_2)$
to the flux without coupling,
i.e.\ we divide the difference by $\coup^2$.
One observes
in \prettyref{fig:plot_flux_p2_increasing_coupling}(b)
that the curves almost coincide.
The remaining small variations
indicate some higher-order contribution of the
small coupling $\coup \ll 1$.

\section{Partial barrier from \threeD{} stable and unstable manifolds of a NHIM}\label{sec:pbmanifold}
\subsection{Motivation}
In \twoD{} maps there is another important partial barrier
besides the partial barrier based on a cantorus.
It arises from
a \zeroD{} hyperbolic fixed (or periodic) point
and its \oneD{} stable and unstable manifolds~\cite{Mei1992,Mei2015}.
One can generalize this partial barrier to
the case of a \fourD{} map
by considering the \twoD{} \NHIM{} related to the hyperbolic fixed point
and its \threeD{} stable and unstable manifolds
\cite{Wig1990, GilEzr1991}.
We want to demonstrate that the flux~$\flux^\fourD{}$ across this partial barrier
composed of \threeD{} stable and unstable manifolds
can be determined
by the flux across appropriate periodic \NHIMs{}.
This approach is similar to \twoD{} maps~\cite{Mei1992},
where the flux across a partial barrier composed
of stable and unstable manifolds of a periodic orbit with frequency $\frac{m}{n}$
can be determined by increasingly higher-order periodic orbits with frequencies
converging to $\frac{m}{n}$.

Furthermore we want to study the local flux
$\flux^\threeD (p_2)$ for this partial barrier.
In particular, we want to test if the \NHIM{} contribution
captures the prominent features of the $p_2$ dependence of the local flux
also for this type of partial barrier,
which is composed of invariant manifolds.

As an aside, we mention that one can avoid the
determination of \NHIMs{}
by constructing corresponding \twoD{} manifolds
by interpolating families of \oneD{} hyperbolic tori
\cite{Hue2020}. This gives approximately the same global and local
flux, but becomes less accurate for increasing coupling,
as the dynamics on the \NHIMs{} has larger resonances and chaotic regions
(see \prettyref{fig:nhim_poincare_increasing_period} in the appendix).

\subsection{Partial barrier and flux}
\begin{figure}
    \includegraphics{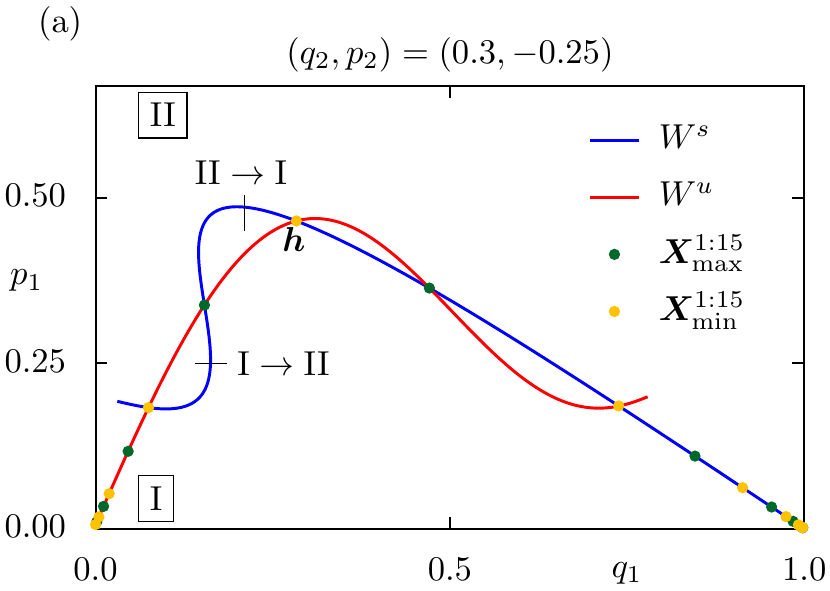}
    \includegraphics{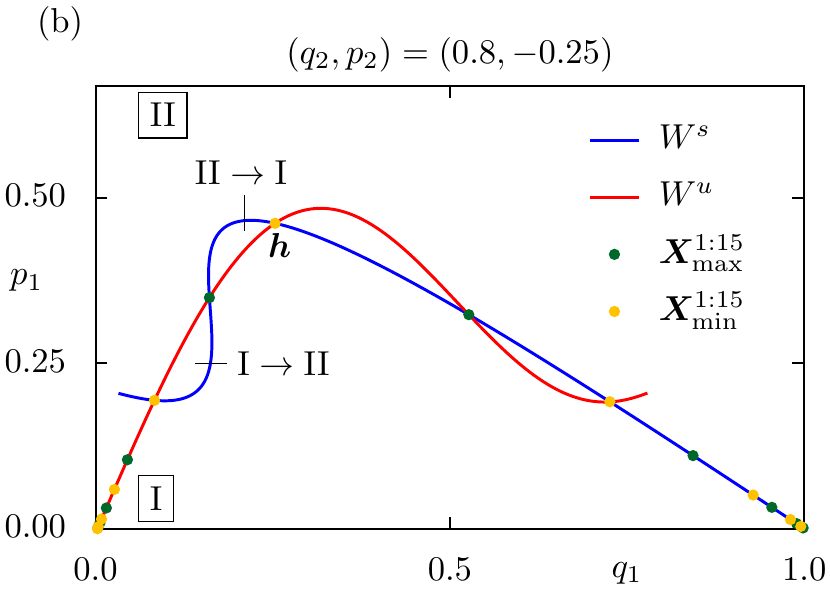}
    \caption{
        Partial barrier composed of segments of
        the \threeD{} stable (\manistable{}) and unstable (\maniunstable{}) manifolds
        of the \twoD{} \NHIM{} \nhimpomnmin{0:1}{}
        merging at the \twoD{} homoclinic manifold $\bm h$
        in a $(q_2, p_2)$ section of phase space
    	for (a) $(q_2, p_2) = (0.3, -0.25)$
		and (b)  $(q_2, p_2) = (0.8, -0.35)$
        for $\coup = 0.05$.
        Exit lobes from region~\regionone{} and from region~\regiontwo{}
        are shown.
        Periodic \NHIMs{} \nhimpomnmax{1:15}{} (dark green points)
        and  \nhimpomnmin{1:15}{} (light orange points) approximate
        the intersections of stable and unstable manifolds.
    }
    \label{fig:plot_manifold_section}
\end{figure}

We consider the \fourD{} map \mappingfourd{} with parameters as in Sec.~\prettyref{subsec:system}.
The \twoD{} \NHIM{} \nhimpomnmin{0:1} corresponds to the
hyperbolic fixed point $\pomnmin{0:1} = (0,0)$ of the \twoD{} map \mappingtwod.
This \NHIM{} has a \threeD{} stable manifold \manistable{}
and a \threeD{} unstable manifold \maniunstable{}.
The manifolds \manistable{} and \maniunstable{}
intersect (infinitely often) in \twoD{} homoclinic manifolds.
This is shown for two $(q_2, p_2)$ sections of the
\fourD{} phase space in \prettyref{fig:plot_manifold_section},
where the \threeD{} manifolds appear as \oneD{} lines
intersecting in \zeroD{} points.

The \threeD{} partial barrier is composed of segments of
\manistable{} and \maniunstable{} ranging from the \NHIM{} \nhimpomnmin{0:1}
to a specific chosen homoclinic manifold $\bm h$,
see \prettyref{fig:plot_manifold_section}.
It separates phase space
into regions~\regionone{}  and~\regiontwo{}.
There is one \fourD{} exit lobe from region~\regionone{} and
one exit lobe from region~\regiontwo{},
each enclosed by segments of the stable and unstable manifolds.
For this exact volume preserving map they lead to the same
global flux~$\flux^\fourD$.
These lobes appear as \twoD{} areas
in the $(q_2, p_2)$ sections of \prettyref{fig:plot_manifold_section}
of varying sizes.
The global flux~$\flux^\fourD$
for $\coup = 0.05$
is determined by integrating
the \twoD{} areas over all  $(q_2, p_2)$ sections,
giving $\flux^\fourD = 0.0064679$ for each lobe.

\subsection{Flux from periodic \NHIMs{}}

\begin{table}
	\begin{tabular}{ c| c | c}
		$m\!:\!n$ & $\flux^{\fourD}(\coup = 0)$ &$\flux^{\fourD}(\coup = 0.05)$   \\
		\hline
		$1\!:\!3$ &   $0.0094004$ & $0.0094771$ \\
		$1\!:\!5$ &   $0.0066441$ & $0.0067190$ \\
		$1\!:\!7$ &   $0.0064130$ & $0.0064891$ \\
		$1\!:\!9$ &   $0.0063935$ & $0.0064697$ \\
		$1\!:\!11$ &  $0.0063918$ & $0.0064680$ \\
		$1\!:\!13$ &  $0.0063917$ & $0.0064679$ \\
		$1\!:\!15$ &  $0.0063917$ & $0.0064679$ \\
	\end{tabular}
	\caption{Flux $\flux^{\fourD}$, \prettyref{eq:fluxfourd},
		across partial barriers based on periodic \NHIMs{}
		\nhimpomn{1:3}{} to \nhimpomn{1:15}{}
		for coupling $\coup = 0$ and $\coup = 0.05$,
		showing fast convergence with increasing period $n$.
	}
	\label{tab:flux_4d_mani}
\end{table}

In \twoD{} maps one can use periodic orbits
to determine the flux across a partial barrier composed
of stable and unstable manifolds.
This is possible
as appropriate periodic orbits approximate the
homoclinic intersections of stable and unstable manifolds.
There is a flux formula based on these (infinitely many) homoclinic intersections~\cite{Mei1992}.
These ideas have been generalized
to arbitrary dimensions~\cite{LomMei2009, Mei2015}.

We want to use periodic \NHIMs{}, introduced in
\prettyref{subsec:periodicnhim},
to determine
the global flux~$\flux^\fourD{}$ across a partial barrier
composed of \threeD{} stable and unstable manifolds.
This has the advantage that it does not require
the numerical determination of \threeD{} stable and unstable manifolds
and their \twoD{} homoclinic intersections.

We use periodic \NHIMs{} \nhimpomnmax{1:n}{} and \nhimpomnmin{1:n}{}
with frequencies $\nu = \frac{1}{n}$ for increasing $n$.
They
appear as \zeroD{} periodic points in the $(q_2, p_2)$ sections of \prettyref{fig:plot_manifold_section},
where they approximate several homoclinic intersections of the stable and unstable manifolds
of the \NHIM{} \nhimpomnmin{0:1}.
The flux~$\flux^\fourD{}$ across periodic \NHIMs{} is determined using \prettyref{eq:fluxfourd}.
It converges for increasing $n$, as shown in \prettyref{tab:flux_4d_mani},
agreeing with the above \fourD{} volume measurement of the exit lobe.
Note that the flux for this type of partial barrier is about $50\%$ larger
than for the cantorus-\NHIM{} for the considered parameters, see
\prettyref{tab:flux_4d},
as is the case for the \twoD{} map \mappingtwod.
This also quantitatively confirms that the cantorus-\NHIM{} provides
the more restrictive transport barrier.

\subsection{Local flux}\label{subsec:local_flux_mani}
\begin{figure}
    \centering
    \includegraphics{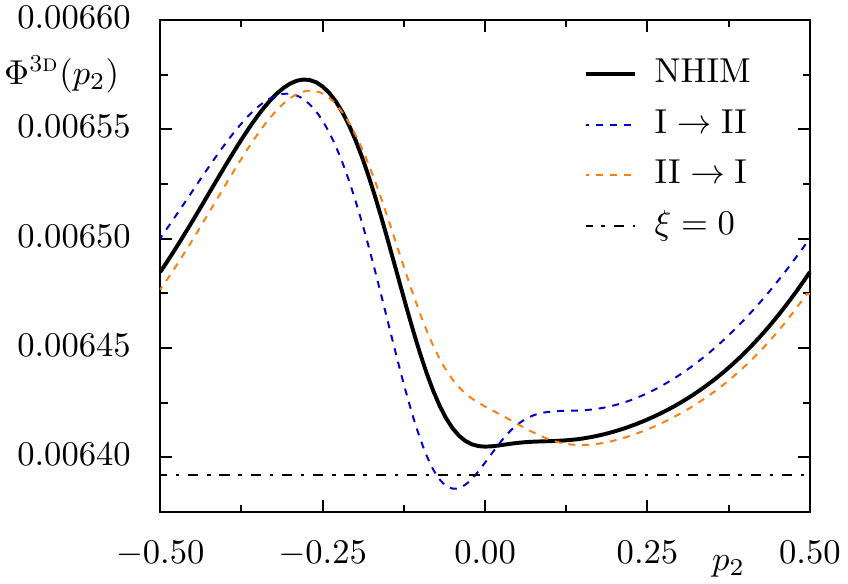}
    \caption{
		Local flux $\flux^\textsc{3d} (p_2)$, \prettyref{eq:fluxthreed},
		for the \threeD{} partial barrier
		composed of segments of
		the \threeD{} stable and unstable manifolds
		of the \twoD{} \NHIM{} \nhimpomnmin{0:1}{}
		in direction $\regionone \rightarrow \regiontwo$
		(dark blue dashed line)
		and $\regiontwo \rightarrow \regionone$
		(light orange dashed line)
		based on the periodic \NHIM{} \nhimpomn{1:15}{}
		for coupling $\coup = 0.05$.
		Comparison to the local
		flux without coupling, $\flux^{\textsc{3d}}(\coup = 0)$
		(black dash-dotted line)
		and to the \NHIM{} contribution,
		$\flux^\threeD_\text{NHIM} (p_2)$,
		\prettyref{eq:nhimcont}
		(black thick line).
    }
    \label{fig:plot_flux_p2_manifold}
\end{figure}

We study the local flux
$\flux^\threeD (p_2)$ for the partial barrier
composed of segments of
the \threeD{} stable and unstable manifolds
of the \twoD{} \NHIM{} \nhimpomnmin{0:1}{}.
To this end, we determine the local flux
from region~$\regionone{}$ to region~$\regiontwo{}$ and vice versa,
by integrating the \twoD{} areas in the $(q_2, p_2)$ sections with respect to $q_2$,
\prettyref{eq:fluxthreed}.
The result for the local flux $\flux^\threeD (p_2)$
is depicted in \prettyref{fig:plot_flux_p2_manifold},
showing the $p_2$-dependence
for this kind of partial barrier and
a weak dependence on the direction showing that there is no
zero net flux in a given $p_2$ section.

We find that the \NHIM{} contribution $\flux^\threeD_\NHIM (p_2)$
captures the prominent features of the $p_2$ dependence of the local flux
also for this type of partial barrier,
see \prettyref{fig:plot_flux_p2_manifold}.
Here the partial barrier is defined by naturally occurring invariant manifolds,
in contrast to \prettyref{sec:localflux}, where
infinitely many partial barriers joining
the \NHIMs{} could be used.
This shows that the \NHIM{} contribution
for the local flux gives the most important contribution,
both for the partial barrier based on the cantorus \NHIM{} and
for partial barriers from \threeD{} stable and
unstable manifolds of a \NHIM{}.

Note that one can study other partial barriers,
by using a different homoclinic manifold than $\bm h$,
where stable and unstable manifolds are joined,
see \prettyref{fig:plot_manifold_section}.
For example, one could use $\mappingfourd(\bm h)$
further to the right or $\mappingfourd^{-1}(\bm h)$ further to the left.
In these cases we find that the deviation of the local flux
from the \NHIM{} contribution is larger, but
also the difference between the local flux in the two directions
is increased~\cite{Fir2020}.

\section{Summary and outlook}\label{sec:outlook}

For \fourD{} symplectic maps we propose a partial barrier
based on the new concept of a cantorus-\NHIM{}.
To determine the flux across such a partial barrier
we establish the relevance of periodic \NHIMs{}
consisting of $n$ \twoD{} manifolds.
These generalize the \zeroD{} periodic orbits, which
are crucial for the approximation of a cantorus in
\twoD{} symplectic maps.
We determine the \fourD{}
flux~$\flux^\fourD$, \prettyref{eq:fluxfourd},
across a partial barrier based on a pair of periodic \NHIMs{}.
The flux can be determined directly from
the \twoD{} manifolds of
the periodic \NHIMs{}.
We check this approach by comparing with explicit volume measurements
of the \fourD{} lobe volumes.
By increasing the periodicity,
while approximating with $\frac{m}{n}$ an irrational frequency,
these periodic \NHIMs{} approximate a cantorus \NHIM{} and
the \fourD{} flux converges.

In the presence of slow Arnold diffusion along a resonance channel
the global \fourD{} flux for
the entire resonance channel
gives limited information.
Therefore we introduce the more relevant local \threeD{} flux,
depending on the position along a resonance channel.
This local \threeD{} flux turns out to depend on the specific choice
of the partial barrier even when based on the same
pair of periodic \NHIMs{}.
The relevant common features
of the dependence on the position along the resonance channel,
however,
are very well described by a \NHIM-dependent contribution
$\flux^\threeD_\text{NHIM} (p_2)$,
\prettyref{eq:nhimcont}.
This allows for describing the local flux from
one resonance channel to another resonance channel
independent of a specific partial barrier and just based
on the \twoD{} manifolds of
a pair of periodic \NHIMs{}.

Finally, we utilize periodic \NHIMs{} to quantify the flux across a
partial barrier composed of stable and unstable manifolds of a \NHIM{},
both for the global \fourD{} flux and the local \threeD{} flux.
This answers a question raised in Ref.~\cite{Mei2015},
\emph{Question IV  (Multidimensional Flux).
Can one use the
flux formulas to compute flux [...]
through resonance zones formed
from \twoD{} normally hyperbolic invariant sets in a \fourD{} symplectic map
[...]?},
in the affirmative.

The present work opens the path for the analysis of more generic
situations in \fourD{} symplectic maps.
Here, we rely on a parameter regime
for which the \NHIMs{} emerge from an uncoupled case.
When increasing the coupling strength
the hyperbolicity on the \NHIMs{} increases and will eventually
lead to a breakup of the \NHIMs{} \cite{GonDroJun2014, TerTodKom2015, Jun2021}.
If just part of the \NHIM{} exists, the
assumptions of \prettyref{sec:pbfluxgen} are not fulfilled
and a global \fourD{} flux can no longer be determined.
However, a local \threeD{} flux may still be well defined.

Furthermore, the partial barriers studied here are
in frequency space all parallel to a frequency axis and
they are in between parallel resonance channels.
Generically, one has transport between resonance channels
that are not parallel to a frequency axis
and that are not parallel to each other,
as observed e.g.\ in Refs.~\cite{MarDavEzr1987, Las1993, LanBaeKet2016, FirLanKetBae2018}.
Still we expect that periodic \NHIMs{} occur and can be used
to determine the local flux.
We expect that such a local flux from one resonance channel to another
will depend much stronger on the position along the resonance channel
than in the examples studied here.
In particular, we expect the local flux to be
small near a predominantly regular region
and to increase towards the chaotic sea.
Understanding this local flux will give new insights in finding
the mechanism of stickiness leading to power-law trapping in higher-dimensional
systems~\cite{LanBaeKet2016}.

Moreover, the partial barriers based on periodic \NHIMs{}
can be directly generalized to higher-dimensional symplectic maps
with weakly coupled degrees of freedom.
In case of an $n$-dimensional map the corresponding
\NHIMs{} consist of $(n-2)$-dimensional manifolds.
An $(n-1)$-dimensional partial barrier based on them will allow for
an $n$-dimensional flux.
The flux can be determined in analogy to
\prettyref{eq:fluxfourd}
from the properties of a pair of periodic \NHIMs{} only.
The ideas presented here should also be relevant and
applicable to higher-dimensional time-continuous systems
and transition state theory in particular.

Finally, consequences of a partial barrier on the
corresponding quantum system are of interest.
It is known from the lower-dimensional case that the flux across a partial barrier
has to be compared to the size of a Planck cell~\cite{KayMeiPer1984b, MicBaeKetStoTom2012, KoeBaeKet2015}:
If the Planck cell is larger than the flux one finds that
quantum mechanically transport is blocked,
while in the opposite case classical transport is mimicked.
The transition is described by a universal scaling curve.
It is an interesting question to find out the relevant scaling
in the higher-dimensional case. Presumably, the ratio of \fourD{} flux to
\fourD{} Planck cell is relevant. But it is not clear, if
this ratio enters the universal scaling curve or its square root,
which describes the ratio per degree of freedom.
This question becomes even more relevant in higher dimensions.
Additionally, it will be worth exploring the quantum mechanical
impact of a strongly varying local \threeD{} flux.

\acknowledgments

We are grateful for discussions with Felix~Fritzsch,
Franziska~H\"ubner,
Christof~Jung,
Jim~Meiss,
Srihari~Keshavamurthy,
and Jonas St\"ober.
Funded by the Deutsche Forschungsgemeinschaft (DFG, German Research Foundation) – 290128388.

\appendix

\section{Dynamics on \textsc{2D} periodic NHIMs}\label{app:dynonnhim}
\begin{figure}
	\centering
	\includegraphics{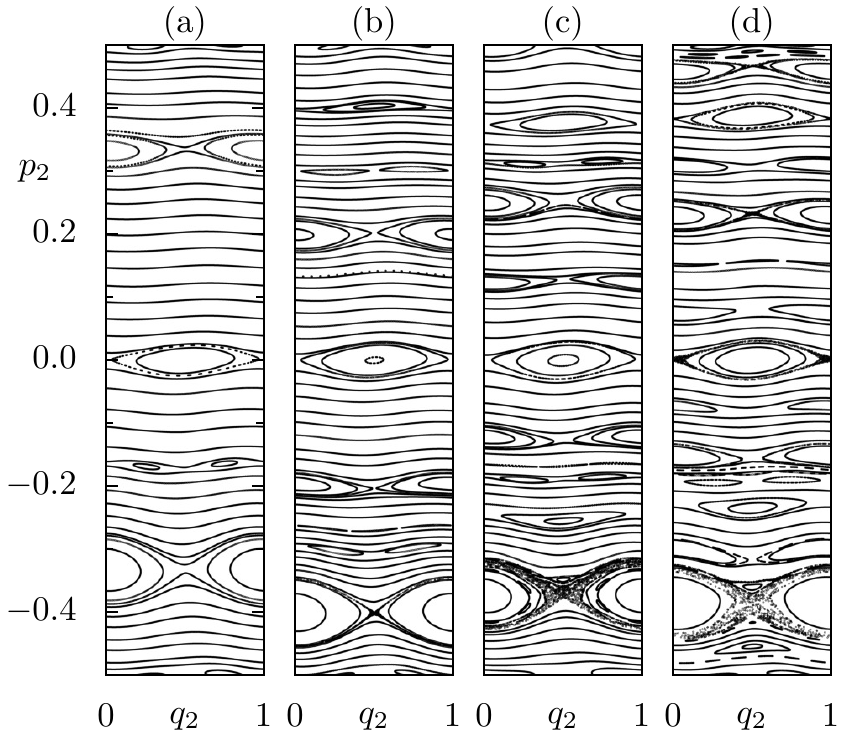}
	\caption{
		Phase-space structures
		on the \twoD{} periodic \NHIM{} \nhimpomnmax{m:n}
		at $\kone = 1.75$, $\ktwo = 0$, $\coup = 0.03$ for
		frequencies $\frac{m}{n} \in \{\frac{1}{3}, \frac{2}{5}, \frac{3}{8}, \frac{5}{13}\}$
		from (a) to (d),
		visualized by a projection of the manifold with $q_1 \approx 0.5$
		onto $(q_2, p_2)$ space.
	}
	\label{fig:nhim_poincare_increasing_period}
\end{figure}

Each of the $n$ manifolds $\nhimpomn{m:n}{i}$ of a periodic \NHIM{} $\nhimpomn{m:n}{}$ is
invariant under the $n$-fold map,
$\mappingfourd^n(\nhimpomn{m:n}{i}) = \nhimpomn{m:n}{i}$,
for $i= 0, \dots, n-1$.
Therefore one can study the dynamics of the $n$-fold \fourD{} map $\mappingfourd^n$
restricted to one of the \twoD{} manifolds $\nhimpomn{m:n}{i}$.
This is visualized in
\prettyref{fig:nhim_poincare_increasing_period}
for the periodic \NHIM{} \nhimpomnmax{m:n}
for increasing periods $n$ with $\frac{m}{n}$ approaching the golden mean.
Such phase-space portraits for \NHIMs{} based on
hyperbolic fixed points (i.e.\ with period $n=1$)
have been studied
in Refs.~\cite{LiShoTodKom2006a,GonDroJun2014, GonJun2015, TerTodKom2015, CanHar2016, GonJun2022}.

Without coupling, $\coup = 0$,
one finds integrable dynamics for $\ktwo=0$,
as shown in \prettyref{fig:ps_product} (right).
A small coupling effectively acts as perturbation
leading to resonances and chaotic layers,
as in a non-integrable \twoD{} symplectic map,
see \prettyref{fig:nhim_poincare_increasing_period}.
With increasing coupling the size of the resonances and
the associated chaotic layers increase~\cite{Fir2020}.

The occurring resonances on the manifold are related to
the frequency $\frac{m}{n}$ of the \NHIM{} \nhimpomn{m:n}{}.
They occur at so-called resonance junctions\cite{Hal1995, Hal1999, GotNoz1999, Got2006, EftHar2013, KarKes2018}
in frequency space, where the resonance line
${n:0:m}$
intersects with the periodic \NHIM{}.
The biggest resonance (at $p_2 = - \frac{m}{n}$)
appears at the crossing with the
${1:1:0}$ resonance channel.
The second biggest resonance (at $p_2 =0$)
occurs at the crossing with the
${0:1:0}$ resonance channel.

We mention that the phase-space portraits of all
manifolds \nhimpomn{m:n}{i} with $i = 0, \dots, n-1$ show qualitatively the same
phase-space structure
with some shift in $q_2$-direction.
The phase-space portraits of  \nhimpomnmin{m:n} (not shown)
do not differ substantially from the ones
shown for \nhimpomnmax{m:n}
in \prettyref{fig:nhim_poincare_increasing_period}.

Numerically,
one has to deal with the instability of the dynamics normal to
the periodic \NHIM{}.
Thus after each iteration of the map \mappingfourd{}, \prettyref{eq:map_stdfourd},
we project the \fourD{} orbit onto the
corresponding manifold \nhimpomn{m:n}{i}.
This is done by using the iterated $(q_2, p_2)$ coordinates
and determining the corresponding $(q_1, p_1)$ coordinates of the \NHIM{}
using the parametrization, \prettyref{eq:nhimpara}.

\section{Flux $\flux^\textsc{2D}$ in $(q_2, p_2)$ section}\label{app:lobearea}

A \threeD{} partial barrier $C$ and its
\threeD{} preimage $\mappingfourd^{-1} (C)$ give rise to
\fourD{} lobe volumes
transported from region~\regionone{} to~\regiontwo{}.
For the computation of this \fourD{} volume
in \prettyref{eq:flux4dint} we use
an integral over $(q_2, p_2)$ sections of phase space.
The objects in such a section appear in the remaining $(q_1, p_1)$ space
with a dimension reduced by two.
Here the partial barrier and its preimage
are \oneD{} lines,
see \prettyref{fig:pbnhimprojpreimage}.
Together they enclose \twoD{} lobe areas
which vary with the considered $(q_2, p_2)$ section.

The sum of the lobe areas in region~\regionone{} gives the
flux~$\flux^\twoD_{\regionone \rightarrow \regiontwo} (q_2, p_2)$ for a chosen
$(q_2, p_2)$ section.
This flux can be directly evaluated from
\prettyref{eq:flux_3}
by considering the $(q_2, p_2)$ sections of the \threeD{} manifolds
$U$ and $S$, giving the \oneD{} manifolds
$U_{q_2, p_2}$ and $S_{q_2, p_2}$, respectively.
Let us remind, that $U$ and $S$ enclose the exit set,
see \prettyref{fig:sketch_lobe} for $n=2$,
and $U$ is a subset of the partial barrier $C$,
while $S$ is a subset of the preimage,
such that $\mappingfourd{}(S)$ is also a subset of $C$.
In contrast to $\mappingfourd{}(S) = U$ used for the
derivation of \prettyref{eq:flux_general},
for the here considered $(q_2, p_2)$ section
we have $\mappingfourd{}(S_{q_2, p_2}) \ne U_{q_2, p_2}$
and the last two integrals in \prettyref{eq:flux_3}
do not cancel.
This leads for the map \mappingfourd{} with the generating
two-form $\gentwoformfourd$,
\prettyref{eq:gentwostd4d}, to
\begin{equation}
\begin{aligned}
\flux^\twoD_{\regionone \rightarrow \regiontwo} (q_2, p_2)
=& \int \limits_{\partial U_{q_2, p_2}}
\frac12 p_1^2 - V(q_1 + p_1, q_2 + p_2)
\\
&+ \int \limits_{U_{q_2, p_2}} p_1 \: \ud q_1
- \int \limits_{\mappingfourd(S_{q_2, p_2})} p_1 \: \ud q_1
.
\end{aligned}
\label{eq:flux_3_section}
\end{equation}
The \zeroD{} boundary points $\partial U_{q_2, p_2}$ are
on the pair of periodic \NHIMs{},
\nhimpomnmin{m:n} and \nhimpomnmax{m:n},
which are
on the left (L) and on the
right (R) end of the \mbox{$i$-th} lobe,
respectively,
\begin{equation}
	\bm x_i^{\text{L/R}}
	= (q_{1, i}^{\text{L/R}}(q_2, p_2),
	p_{1, i}^{\text{L/R}}(q_2, p_2),
	q_2, p_2)
	,
	\label{eq:x_NHIM}
\end{equation}
for $i = 0, \dots, n-1$.
Here the $(q_1, p_1)$ coordinates
depend on the considered $(q_2, p_2)$ section
according to the parametrization of the \NHIMs,
\prettyref{eq:nhimpara}.

For the iterated set $\mappingfourd(S_{q_2, p_2})$
we use the mapping, \prettyref{eq:map_stdfourd}.
For a \threeD{} partial barrier,
$p_1^\text{pb} (q_1, q_2, p_2)$,
we find for the flux~$\flux^\twoD$ in a $(q_2, p_2)$ section
the explicit expression
\begin{equation}
\begin{aligned}
	\flux^\twoD_{\regionone \rightarrow \regiontwo} (q_2, p_2)
	=
	&\sum_{i=0}^{n-1} \: \left[ \frac{1}{2} p_1^2 - V(q_1 + p_1, q_2 + p_2 )\right]^{\bm x_i^{\text{R}}}_{\bm x_i^{\text{L}}}\\
	&+ \sum_{i=0}^{n-1} \:  \int \limits_{q_{1, i}^{\text{L}}}^{q_{1, i}^{\text{R}}} p_1^\text{pb} (q_1, q_2, p_2) \: \ud q_1 \\
	&- \sum_{i=0}^{n-1} \:  \int \limits_{q_{1, i}^{\text{L}} + p_{1, i}^{\text{L}}}^{q_{1, i}^{\text{R}} + p_{1, i}^{\text{R}}} p_1^\text{pb} ( q_1, q_2', p_2') \: \ud q_1 ,
\end{aligned}
\label{eq:lobe_area}
\end{equation}
with $q_2' = q_2 + p_2$ and $p_2' = p_2 - V_2 (q_1, q_2 + p_2)$
according to \prettyref{eq:map_stdfourd}.

In order to compute the flux $\flux^\twoD_{\regiontwo \rightarrow \regionone}$
for the opposite transport direction,
the other set of lobe areas have to be considered.
This is done by interchanging the role of the left and right \NHIM{}
in \prettyref{eq:lobe_area}
and increasing the index $i$ for the right \NHIM{} by one.
Additionally, the overall sign has to be changed to get
a positive flux
$\flux^\twoD_{\regiontwo \rightarrow \regionone}$.

\section{\NHIM{} contribution to local flux~$\flux^\textsc{3D}(p2)$} \label{app:nhimcont}

\begin{figure}
	\includegraphics{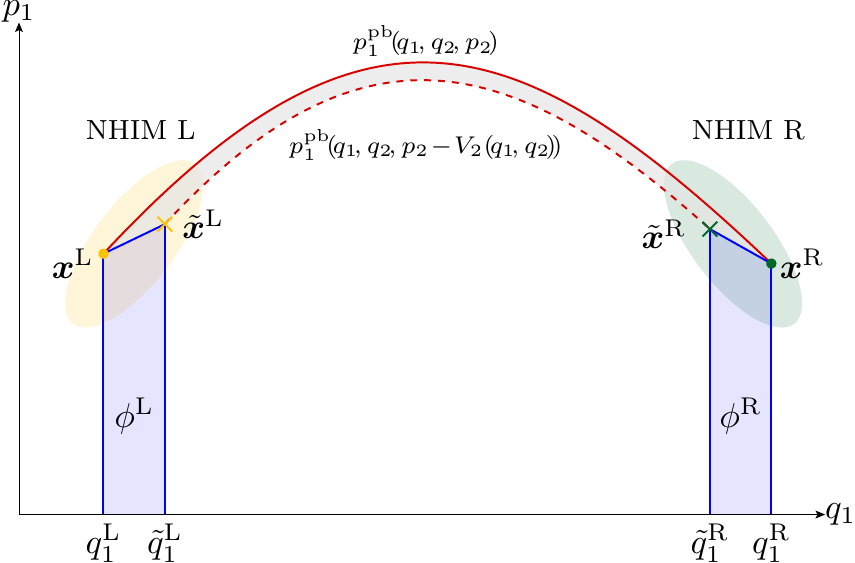}
	\caption{
		Visualization of integrands appearing in
		\prettyref{eq:flux_threed_simple}.
		Partial barrier $C$ for fixed $(q_2, p_2)$
		in $(q_1, p_1)$ space,
		$p_1^\text{pb} (q_1, q_2, p_2)$
		(red solid line),
		between endpoints
		$\bm x^{\text{L/R}}$
		(light orange and dark green point)
		on left and right \NHIM{}
		(ellipsoidal shaded areas).
		Partial barrier $C$ at different $p_2$ coordinate,
		$p_1^\text{pb} (q_1, q_2, p_2 - V_2 (q_1, q_2))$
		(red dashed line),
		between endpoints
		$\tilde{\bm x}^{\text{L/R}}$
		(light orange and dark green cross).
		The difference of areas under the curves is shaded
		and consists of the trapezoidal
		\NHIM-dependent
		areas $\phi^{\text{L/R}}$ (blue)
		and the barrier-specific area (gray).
	}
	\label{fig:sketch_deriv}
\end{figure}

We derive the splitting of the local flux $\flux^\threeD (p_2)$, \prettyref{eq:flux_threed_splitting},
into the \NHIM{} contribution
$\flux^\threeD_\text{NHIM} (p_2)$
and the barrier-specific contribution
$\delta_\text{pb}(p_2)$.
Starting point is \prettyref{eq:fluxthreed},
where the flux $\flux^\twoD(q_2, p_2)$, \prettyref{eq:lobe_area},
is inserted,
\begin{equation}
\begin{split}
&\flux^\threeD{} (p_2) = \int \limits_{0}^{1} \ud q_2  \\
&\quad
\begin{aligned}[t]
\times \Bigg( \;  & \sum_{i=0}^{n-1} \: \left[ \frac{1}{2} p_1^2 - V(q_1 + p_1, q_2 + p_2 )\right]^{\bm x_i^{\text{R}}}_{\bm x_i^{\text{L}}}\\
& + \sum_{i=0}^{n-1} \: \int \limits_{q_{1, i}^{\text{L}}}^{q_{1, i}^{\text{R}}} p_1^\text{pb} (q_1, q_2, p_2) \: \ud q_1\\
{}& - \sum_{i=0}^{n-1} \: \int \limits_{\tilde{q}^{\text{L}}_{1, i}}^{\tilde{q}^{\text{R}}_{1, i} }
p_1^\text{pb} \big(q_1, q_2, p_2 - V_2 (q_1, q_2) \big) \: \ud q_1 \Bigg) .
\end{aligned}
\end{split}
\label{eq:flux_threed_simple}
\end{equation}
In the last line the integration variable $q_2$ is
changed to $q_2' = q_2 + p_2$ and the periodicity in
$q_2$ is used
with the prime omitted afterwards.
This change of variables has the advantage that now
both integrands of the last two lines have the same $q_2$ coordinate
and for the considered small coupling  almost the same $p_2$ coordinate,
such that the difference of the integrals is small,
as used below.

Moreover, in the last line of
\prettyref{eq:flux_threed_simple} we use
the coordinates $\tilde{q}^{\text{L/R}}_{1, i}$ of the points
\begin{equation}
	\tilde{\bm x}_i^{\text{L/R}}
	= \mappingfourd\big(\bm x_{i-1}^{\text{L/R}}(q_2 - p_2, p_2)\big)
	,
	\label{eq:x_tilde_NHIM}
\end{equation}
marked by a tilde, which are the iterates of the
boundary points $\partial S_{q_2-p_2, p_2} = \partial U_{q_2-p_2, p_2}$ on the \NHIMs,
where the index $i$ of the \NHIM{} manifold is changed by one,
such that the points
$\tilde{\bm x}_i^{\text{L/R}}$
are near the corresponding points
$\bm x_i^{\text{L/R}}$,
\prettyref{eq:x_NHIM}.
In terms of coordinates we have
\begin{equation}
	\begin{aligned}
		\tilde{\bm x}_i^{\text{L/R}}
		&= \big(\tilde{q}_{1, i}^{\text{L/R}},
		\tilde{p}_{1, i}^{\text{L/R}},
		q_2, p_2- V_2 (\tilde{q}^{\text{L/R}}_{1, i}, q_2) \big)
		,
	\end{aligned}
\end{equation}
with the $\tilde{q}^{\text{L/R}}_{1, i}$ and
$\tilde{p}^{\text{L/R}}_{1, i}$ determined according to
\prettyref{eq:x_tilde_NHIM}
using \prettyref{eq:map_stdfourd},
\begin{equation}
\begin{aligned}	\tilde{q}^{\text{L/R}}_{1, i}
	&= q_{1, i-1}^{\text{L/R}} (q_2 - p_2, p_2) + p_{1, i-1}^{\text{L/R}} (q_2 - p_2, p_2) ,
	\\
	\tilde{p}^{\text{L/R}}_{1, i}
	&= p_{1, i-1}^{\text{L/R}} (q_2 - p_2, p_2)
	- V_1(\tilde{q}^{\text{L/R}}_{i, 1}, q_2)
	.
	\label{eq:path_abbrev}
\end{aligned}
\end{equation}
These points $\tilde{\bm x}_i^{\text{L/R}}$
are by construction
on the left and the right \NHIM{}, respectively,
but in contrast to
the points $\bm x_i^{\text{L/R}}$,
\prettyref{eq:x_NHIM},
have a slightly different $p_2$ coordinate.

The first term
in \prettyref{eq:flux_threed_simple}
depends on properties of the \NHIMs{} only.
It is similar to
\prettyref{eq:fluxfourd},
up to the missing integration with respect to $p_2$.
One expects that this term
describes the dominant features of the local flux.
However, this term alone is insufficient to describe the local flux
as we find that it varies
as a function of $p_2$ on a roughly 10 times larger scale (not shown)
compared to the local flux~$\flux^\threeD (p_2)$
in \prettyref{fig:flux_p2_compare}.

Thus also the last two lines of
\prettyref{eq:flux_threed_simple}
have to  be considered to find the relevant contribution of
the local flux~$\flux^\threeD (p_2)$,
which depends on the \NHIMs{} only.
A sketch of these integrals
(without $q_2$ integration and without summation over $i$)
is shown in \prettyref{fig:sketch_deriv}
in the $(q_1, p_1)$ space for fixed $(q_2, p_2)$.
Specifically, both integrands and their integration limits are
sketched.
For the local flux,  \prettyref{eq:flux_threed_simple},
the small difference of these integrals is relevant,
shown by transparently colored areas.

There are two blue colored parts
of this area, which depend on the \NHIMs{} only,
denoted as
$\phi_{i}^{\text{L/R}}$.
These contributions are within the intervals
$[q_{1, i}^{\text{L}}, \tilde{q}_{1, i}^{\text{L}}]$
and
$[\tilde{q}_{1, i}^{\text{R}}, q_{1, i}^{\text{R}}]$, respectively,
and originate from the different integration limits
in \prettyref{eq:flux_threed_simple}.
For the upper boundary of the areas
$\phi_{i}^{\text{L/R}}$
we choose a straight line
connecting the points
$\bm x_i^{\text{L/R}}$
and
$\tilde{\bm x}_i^{\text{L/R}}$
on the left and the right \NHIM{}, respectively,
projected onto $(q_1, p_1)$ space
(other choices are discussed below).
The trapezoidal areas $\phi_{i}^{\text{L/R}}$ are given by
\begin{equation}
\phi_{i}^{\text{L/R}} =
\left( q_{1, i}^{\text{L/R}} - \tilde{q}_{1, i}^{\text{L/R}} \right)
\; \frac{p_{1, i}^{\text{L/R}}  + \tilde{p}_{1, i}^{\text{L/R}}}{2}
\;
.
\label{eq:trapez}
\end{equation}
We include these two \NHIM{}-dependent areas
to define the \NHIM{} contribution of the local flux,
\prettyref{eq:flux_threed_simple},
\begin{equation}
\begin{split}
& \flux^\threeD_\text{NHIM} (p_2) =  \\
&
\quad
\begin{aligned}[t]
\int \limits_{0}^{1} \ud q_2
\Bigg( \; & \sum_{i=0}^{n-1} \: \left[ \frac{1}{2} p_1^2 - V(q_1 + p_1, q_2 + p_2 )\right]^{\bm x_i^{\text{R}}(q_2, p_2)}_{\bm x_i^{\text{L}}(q_2, p_2)}\\
&+
\sum_{i=0}^{n-1} \left[ \phi_{i}^{\text{R}}(q_2, p_2)
						- \phi_{i}^{\text{L}}(q_2, p_2)
				\right] \Bigg) .
\end{aligned}
\label{eq:nhimcont}
\end{split}
\end{equation}
The remaining term,
$\delta_\text{pb}(p_2) := \flux^\threeD (p_2) - \flux^\threeD_\text{NHIM} (p_2)$,
describes the deviations from the \NHIM{} contribution,
i.e.\ the additional $p_2$ dependence
specific for the considered partial barrier.
It is related to the gray colored area in \prettyref{fig:sketch_deriv}.

We mention as an aside,
that the straight connection
between the points
$\bm x_i^{\text{L/R}}$
and
$\tilde{\bm x}_i^{\text{L/R}}$
used above is not part of the \NHIMs{}.
Thus another conceptual interesting way is to connect the points
by a path which is entirely on the \NHIMs{} \cite{Fir2020}.
Since the results do not change significantly,
this is not further discussed here.


\end{document}